%% file: main.tex
\documentclass[conference]{IEEEtran}
\IEEEoverridecommandlockouts

\usepackage{cite}
\usepackage{amsmath,amssymb,amsfonts}
\usepackage{algorithmic}
\usepackage{graphicx}
\usepackage{textcomp}
\usepackage{xcolor}
\def\BibTeX{{\rm B\kern-.05em{\sc i\kern-.025em b}\kern-.08em
    T\kern-.1667em\lower.7ex\hbox{E}\kern-.125emX}}

\usepackage{fancyhdr}
\usepackage{xurl}
\usepackage[hidelinks]{hyperref}
\usepackage[ruled,vlined]{algorithm2e}
\usepackage{chngcntr}
\usepackage{multicol}
\usepackage{booktabs}
\usepackage{csquotes}

\usepackage[utf8]{inputenc}
\usepackage{quoting}
\usepackage{ragged2e}
\usepackage{cleveref}
\usepackage{soul} 
\usepackage{subcaption}
\crefformat{section}{\S#2#1#3} 
\crefformat{subsection}{\S#2#1#3}
\crefformat{subsubsection}{\S#2#1#3}
\usepackage{tcolorbox}
\usepackage{environ}
\usepackage{xcolor}
\usepackage[tikz]{bclogo}
\usepackage{tikz}
\usetikzlibrary{calc}
\usepackage{lipsum}

\usepackage{tikz}
\definecolor{lightmintbg}{rgb}{1,1,1}
\newcommand*\circleint[1]{\tikz[baseline=(char.base)]{
            \node[shape=circle,draw,inner sep=1pt, fill=lightmintbg] (char) {\footnotesize #1};}}

\newcommand{\sysnamex}{DIKTAMO} 
\newcommand{\sysname}{DIKTAMO~} 
\newcommand{\sysnamebaselinex}{DIKTAMO-baseline} 
\newcommand{\sysnamebaseline}{DIKTAMO-baseline~} 
\newcommand{\sysparallel}{DIKTAMO-parallel~} 
\newcommand{\sysparallelx}{DIKTAMO-parallel} 
\newcommand{\sysproactive}{DIKTAMO-proactive~}
\newcommand{\sysproactivex}{DIKTAMO-proactive} 
\newcommand{\sysbaseline}{DIKTAMO-baseline~}
\newcommand{\sysbaselinex}{DIKTAMO-baseline}
 
\newcommand{\DIKTAMOTitle}{DIKTAMO: Extending CXL for Resilience\\ to CPU Failures}

\begin{document}

\title{\DIKTAMOTitle{}}

\def\confcameraready{} 
\newcommand{\confpubid}{0000--0000/00\$00.00}

\newcommand\confauthors{
Antonis Psistakis,
Burak Ocalan,
Chloe Alverti$^{\mathsection}$,
Fabien Chaix$^{\diamond}$,
Ramnatthan Alagappan, and
Josep Torrellas
}

\newcommand\confaffiliation{
University of Illinois Urbana-Champaign, USA
\hspace{0.3cm}
$^{\mathsection}$National Technical University of Athens, Greece
\hspace{0.3cm}
$^{\diamond}$FORTH, Greece
}

\newcommand\confemail{
\{psistaki, bocalan, ramn, torrella\}@illinois.edu,
xalverti@cslab.ece.ntua.gr,
chaix@ics.forth.gr
}

\author{
  \ifdefined\confcameraready
    \IEEEauthorblockN{\confauthors{}}
    \IEEEauthorblockA{
      \confaffiliation{} \\
      \confemail{}
    }
  \else
    \IEEEauthorblockN{\normalsize{Conference \confyear{} Submission
      \textbf{\confsubmissionnumber}} \\
      \IEEEauthorblockA{
        Confidential Draft \\
        Do NOT Distribute!!\vspace{-0.3cm}
      }
    }
  \fi
}

\maketitle
\thispagestyle{plain}
\pagestyle{plain}


\input{Sections/abstract}


\input{Sections/introduction}
\input{Sections/background}
\input{Sections/overview}
\input{Sections/diktamo-design}
\input{Sections/diktamo-failures}
\input{Sections/methodology}
\input{Sections/evaluation}
\input{Sections/relatedWork}
\input{Sections/conclusion}

\section*{Acknowledgment}

This work was supported in part by NSF under grants CCF
2107470 and CCF 2316233; and by ACE, one of the seven
centers in JUMP 2.0, a Semiconductor Research Corporation
(SRC) program sponsored by DARPA.

\bibliographystyle{IEEEtranS}
\bibliography{refs}

\end{document}

%% file: Sections/abstract.tex
\begin{abstract}

Compute Express Link (CXL) 3.0 and beyond allows the compute nodes of  a cluster to 
share data with hardware cache coherence and at the granularity of a cache
line. This enables shared-memory semantics for distributed computing, but
introduces a new resilience challenge: a node failure leads to the loss of 
the dirty data in its caches, corrupting application state. 
Sadly, the CXL specification does not consider processor failures.
Moreover, when a component fails, the specification tries to 
isolate it and continue application execution; there is no attempt to bring
the application to a consistent state---a step required to recover a shared-memory program.
 
To address these limitations, this paper extends   CXL 
 to be resilient to node failures, and to
correctly recover the application after
node failures. We call the system  {\em \sysnamex}.
To survive node failures, \sysname augments the
coherence transaction of a write with messages that propagate
the update to a small set of other nodes (i.e.,  {\em Replicas}).  
Replicas save the update in a local hardware Logging Unit.
Replication ensures resilience to node failures. Then, at regular intervals, the Logging Units dump the compressed updates to memory.  
After a node failure, recovery involves using the logs 
to bring the directory and memory to a correct state.
Our evaluation with 16 4-core nodes shows that \sysname 
enables fault-tolerant execution
with 30\% slowdown with 3 replicas (or 27\% with 2 replicas)
over a platform without fault-tolerance support. 
\sysname is 2.82x faster than 
ensuring fault tolerance by using a 
write-through protocol.

\end{abstract}

%% file: Sections/introduction.tex
\section{Introduction}\label{sec:intro}

The  latest versions of the Compute Express Link (CXL) 
standard~\cite{cxl-micro23, intro-to-CXL-acm24, cxlspec},  
including CXL 3.0 and beyond, introduce hardware support for 
distributed compute nodes (CNs) to access shared-data stored in Memory Nodes (MNs) with hardware-enforced cache coherence   at the granularity of a cache line.
This  capability blurs the line between shared-memory multiprocessors and distributed clusters, 
raising questions about how distributed applications can benefit from shared-memory programming semantics~\cite{polardb-cxl,tigon-osdi25, rpccool, rpcstoica, hydrarpc-atc24, sosp-cxl}.
However, this same capability complicates a common challenge in distributed computing: 
tolerance to faults.
Traditionally, distributed applications such as databases and key-value stores share data at coarse granularities (e.g., records) and handle faults with relatively high overhead software 
techniques such as data replication or logging.
With CXL, a different approach is needed, as data sharing occurs transparently in hardware and at cache-line granularity.

The CXL specification describes support for reliability, availability, and serviceability
(RAS)~\cite{cxlspec-ras-overview21, ocp-rasapi23,cxlspec}. While the support is extensive,
it has two limitations. First, 
it does not consider processor (i.e., host) failures;
it only considers link, switch, and device (e.g., memory) failures.
The second limitation is that,
when a component fails, the specification tries to isolate it.
There
is no attempt to recover the workload to a consistent state.

In reality, surviving processor failures is crucial, as CPUs
 are responsible for many 
failures in datacenters~\cite{cpuFails-springer26}. 
Moreover, when a node participating in
a shared-memory program fails, the dirty data in its caches is lost. Hence, simply isolating the node and
continuing execution results in using inconsistent application state.
Correct execution requires first the recovery of the application state.

To address these limitations, this paper proposes an extension
to CXL to:
1) be resilient to the failure of CNs and 2) correctly recover the application after
CN failure. We call the system  {\em \sysnamex}.
\sysnamex's extensions are mostly focused on the CXL 
Transaction Layer, and minimally affect the CXL Link Layer.

To handle the failure of CNs, \sysname augments the coherence transaction of a write  with
messages that propagate the update to a small set of 
other CNs (i.e., the {\em Replicas}). The replicas  
save the update in a local hardware {\em Logging Unit}. 
Such replication ensures resilience to CN failures.
The Logging Units compress the data in their logs and,
at regular intervals,
dump it to the memory of Memory Nodes (MNs), which is resilient. 

CN failure is detected by minimally extending the CXL Link Layer.
Recovery is a distributed software-driven process that involves reading the directory state, 
reading the logs in the Logging Units, and updating the directory and memory state. The recovery
restores a consistent  application state, and enables 
forward progress of the application on the healthy nodes.

We evaluate \sysname using simulations of a cluster of 16 CNs of 4 cores each, and 16 MNs  interconnected via a CXL switch. We run parallel engineering applications and the YCSB key-value store. We find that 
\sysname enables fault-tolerant execution with  30\% slowdown with three replicas (or 27\% with two replicas) over the same platform without fault-tolerance support. Further, \sysname is 
2.82x  
faster than
ensuring fault tolerance by using
write-through writes to MNs.
We also evaluate the impact of various aspects of \sysnamex.

The contributions of this work are:

\noindent $\bullet$ The \sysname   architecture that extends  CXL   to make it 
resilient to processor failures.

\noindent $\bullet$ The \sysname cache
coherence protocol, which extends a traditional MESI protocol with data replication transactions.

\noindent $\bullet$ The \sysname  
recovery scheme.

\noindent $\bullet$ 
An evaluation of \sysname in a simulated CXL cluster.

%% file: Sections/background.tex
\section{Background and Motivation}\label{sec:background}

\begin{figure}[t]
    \centering
    \begin{subfigure}[b]{\linewidth}
        \centering
        \includegraphics[width=1\linewidth]{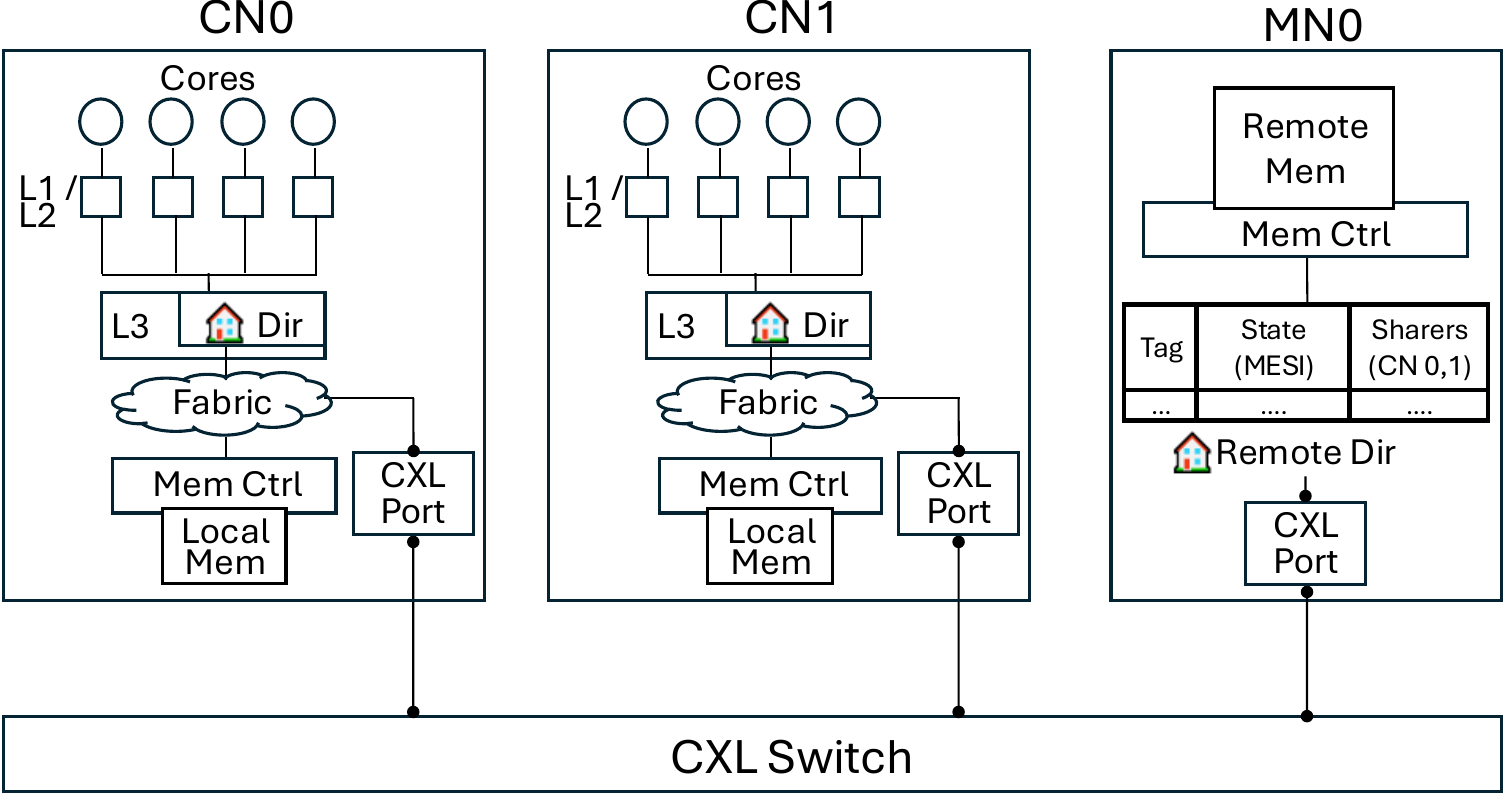} 
    \end{subfigure}
    \caption{CXL-enabled distributed shared-memory system.} 
    \label{fig:background:cxl}
\end{figure}

\subsection{CXL Distributed Shared Memory}
\label{sec:bck:cxl}

Compute Express Link (CXL)~\cite{cxl-micro23, intro-to-CXL-acm24, cxlspec} is an emerging 
open standard high-performance interconnect. The latest versions of its \textit{CXL.mem} protocol (CXL 3.0 and beyond) enable the compute nodes (CNs) of a cluster to directly access data in remote memory nodes (MNs) in a 
byte-addressable and cache-coherent manner.
This unique capability enables CXL-enabled Distributed Shared Memory (CXL-DSM) systems, where coherent data sharing across nodes is supported transparently  by  hardware, and at  cache-line granularity. 

Figure~\ref{fig:background:cxl} shows such a system with two CNs,
one MN, and a CXL switch interconnecting them. 
In general, there are  multiple MNs.
The CNs have their local (i.e., private) memory hierarchies, and can access the
remote memory in the MNs through the CXL switch.
Physical addresses in the local memories are called {\em Local},
while those in the MNs are called {\em Remote}.
On an LLC miss, routing hardware 
identifies if the miss is local or remote based on its target physical address. Remote misses are routed to the CXL port and access the corresponding remote directory in the MNs.
Cores in a CN use write-back caching for both local and remote addresses.
Data coherence is enforced with a hierarchical directory supporting a
conventional cache coherence protocol like MESI:
each CN has a local directory that enforces 
cache coherence locally; the remote directory distributed across  the MNs leverages CXL to
keep the remote addresses
cached in the CNs coherent. 

CXL is organized in three main protocol layers: a) the \textit{Transaction Layer}, which defines the 
protocol transaction types and their ordering rules,
b) the \textit{Link Layer}, which ensures reliable transmission of the transaction packets, 
and c) the \textit{Physical Layer}, which transmits data over the underlying interconnect.

\vspace{-2mm}
\subsection{CXL Reliability}
\label{sec:background:reliability} 
In CXL-DSM clusters, faulty hardware can crash the execution of distributed
shared-memory applications~\cite{apta-dsn23,polardb-cxl,sosp-cxl}. Thus, the CXL specification adds 
support for reliability, availability, and serviceability (RAS)~\cite{cxlspec-ras-overview21, ocp-rasapi23,cxlspec}. 
It considers link, switch, and device (e.g., memory) failures. 
It detects link 
failures via timeouts, and retries packet transmissions to handle them. 
To detect data corruption, it relies on established techniques such as parity.

To contain errors, the CXL specification defines three complementary mechanisms. One is 
\emph{Data Poisoning}, which allows devices to tag messages with corrupted data.
A second one is \emph{Isolation}, which allows the protocol to mark a faulty device and mask traffic from/to the device, keeping the rest of the system operational.  The final one is \emph{Viral Handling}  which, when a fatal  error is detected,  sets a Viral\_Status bit, propagating a 
“viral” state 
across all connected ports and tagging all traffic across ports.
The specification~\cite{cxlspec} also introduces hardware commands to reset (re-initialize)  CXL devices into a clean state after a viral handling.

This CXL RAS model has two limitations. The first one is that
it does not consider processor (i.e., host) failures. 
In reality, handling processor failures is crucial, as CPUs
 are responsible for many  
failures in datacenters~\cite{cpuFails-springer26}. 
When a  host  
fails, the dirty data it stores in its cache hierarchy is permanently lost~\cite{barely}. 

The second limitation is that the 
CXL RAS model focuses only on  containing errors. When a component fails, the specification tries to isolate it.
There
is no attempt to recover lost or corrupted data and  bring the workload to a consistent 
state~\cite{barely}.
In reality, when a node participating in
a shared-memory program fails and its dirty cached data is lost, continuing execution with an
inconsistent state is not acceptable. The workload state must
be corrected.

A possible approach to avoid data loss on a CN failure is to always
use write-through writes for remote physical addresses, 
propagating all updates to 
MNs---while assuming that the MN memory is resilient~\cite{apta-dsn23, polardb-cxl}. However, 
this approach can introduce substantial performance overheads, especially in systems that
support the TSO  
memory consistency model~\cite{x86TSO-cacm10,sparcTSO-94}. In TSO, writes from the same core to 
different addresses
need to be fully serialized. Hence, an update from a core can only be written through to an MN after the previous update from the core
has completed its round trip to an MN (i.e., request and acknowledgment).  

\begin{figure}[t]
    \centering
\begin{subfigure}[t]{\linewidth}
    \centering
   \includegraphics[width=\linewidth]{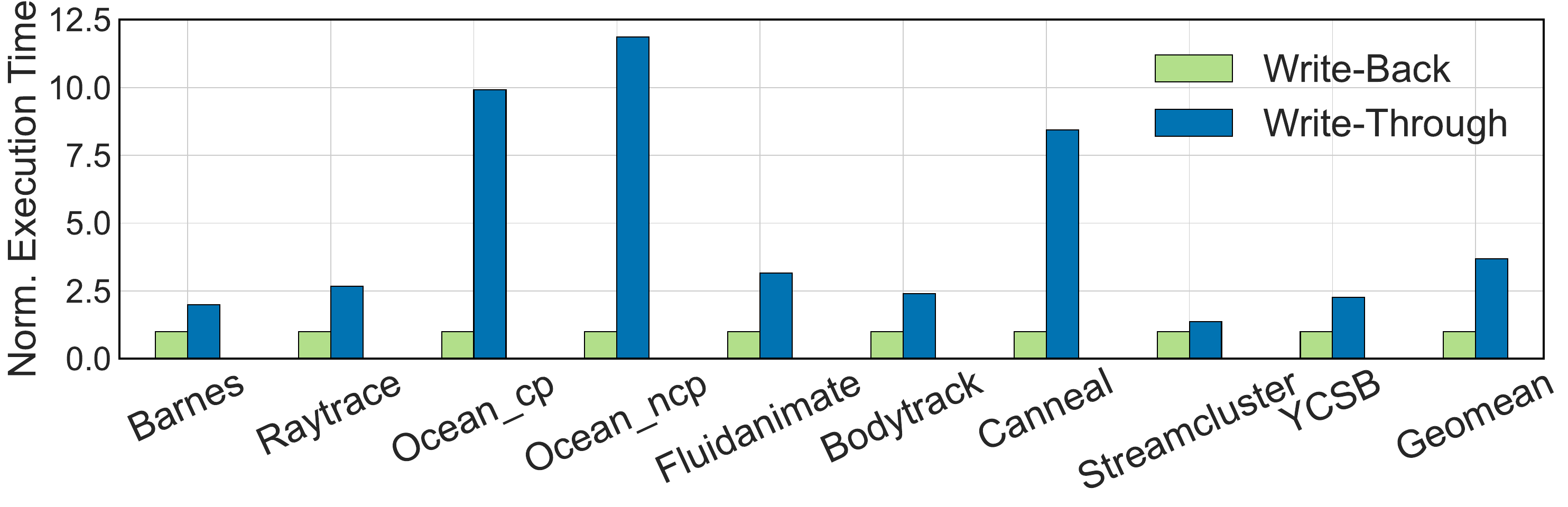} 
\end{subfigure}
    \vspace{-6mm}
    \caption{Application execution time  using a write-back or a write-through 
 protocol. The bars are normalized to the former.} 
   \label{fig:motiv:wbvswt}
\end{figure}

As an example, Figure~\ref{fig:motiv:wbvswt} considers a CXL-DSM
cluster of 16 CNs
of 4 cores each, and 16 MNs interconnected via a CXL switch, running several 64-threaded
shared-memory parallel applications.
The architecture and applications are described in Section~\ref{sec:methodology}. The figure shows
the applications' execution times using  either the MESI  write-back protocol
(WB) or a write through (WT) protocol. The first environment
is not resilient, as dirty cached data can be lost on a CN failure.
The second environment is resilient because, like in 
previous literature~\cite{apta-dsn23, polardb-cxl}, we assume that 
MN memory is fault tolerant and that, as soon as a
write updates an MN’s DRAM memory, the update
cannot be lost. MN memory can be made fault-tolerant through techniques like
replication, persistency, or the use of  more expensive technologies.
How this is done
is not relevant to our work. 

The bars in the figure are normalized to the WB execution time.
We see that WT adds substantial overhead. On average, the execution time of
WT is 3.86x higher than WB's. A major reason for WT's longer execution time  is that,
under TSO, multiple write-through updates
from a core cannot overlap with each other.

As we seek to increase the resilience of CXL, we can get inspiration from existing techniques that
distributed systems use, such as replication of data records across nodes to survive node 
loss~\cite{minos-hpca24, farm-nsdi14, par-atc19, ramcloud-sosp11}, or data logging to recover old
data versions~\cite{singlenode-osdi14, adaptivelogging-sigmod16}. However, these techniques are
typically implemented in software and operate at coarse granularity. In CXL-DSM, where data coherence is maintained at cache-line level  and is enforced in hardware,
we need solutions that are hardware-based, fine-grained, and low-overhead. In 
addition, resilient architectures that use 
CXL memory~\cite{tigon-osdi25,apta-dsn23} 
typically support transactions, which present 
different tradeoffs than our target general-purpose environment. Discussion on 
related work is presented in Section~\ref{sec:relatedWork}.

%% file: Sections/overview.tex
\section{Overview of \sysname}
\label{sec:overview}

In this paper, we address the two limitations of the CXL RAS model described above
by extending  CXL   to handle the  failure of a  CN 
and to correctly recover the application state after a CN failure. We call the proposal {\em \sysnamex}.
\sysnamex's extensions are mostly focused on the CXL 
Transaction Layer and minimally affect the CXL Link Layer.

\vspace{-2mm}
\subsection{Overall Operation}\label{sec:diktamo:over} 

To accomplish its goals, \sysnamex: 1) adds extra functionality to 
ordinary execution, 2) extends error detection, and 3) provides error recovery. In this section, we overview each of
these three components.

\vspace{1mm}
\noindent {\bf 1. Extensions to Ordinary Execution.} \sysname augments the CXL  Transaction Layer for writes  that target the shared  CXL memory (i.e., remote writes). Any such write 
performed by a core in a  CN  is augmented with messages that propagate the update to a small set of 
other CNs (i.e., the {\em Replicas}). 
The replicas  
save the update in a local hardware {\em Logging Unit}. 
The original write cannot commit until the Logging Units in all the replicas are updated. However, crucially,
multiple writes from the same core can overlap their replica update.
The Logging Units compress the data they receive and,
at regular intervals,
dump it  to the memory of  MNs. 
As we describe in Section~\ref{subsec:failure-model}, we consider that the MNs are 
safe from faults.

\begin{figure}[t]
    \centering
    \begin{subfigure}[b]{\linewidth}
        \centering
        \includegraphics[width=\linewidth]{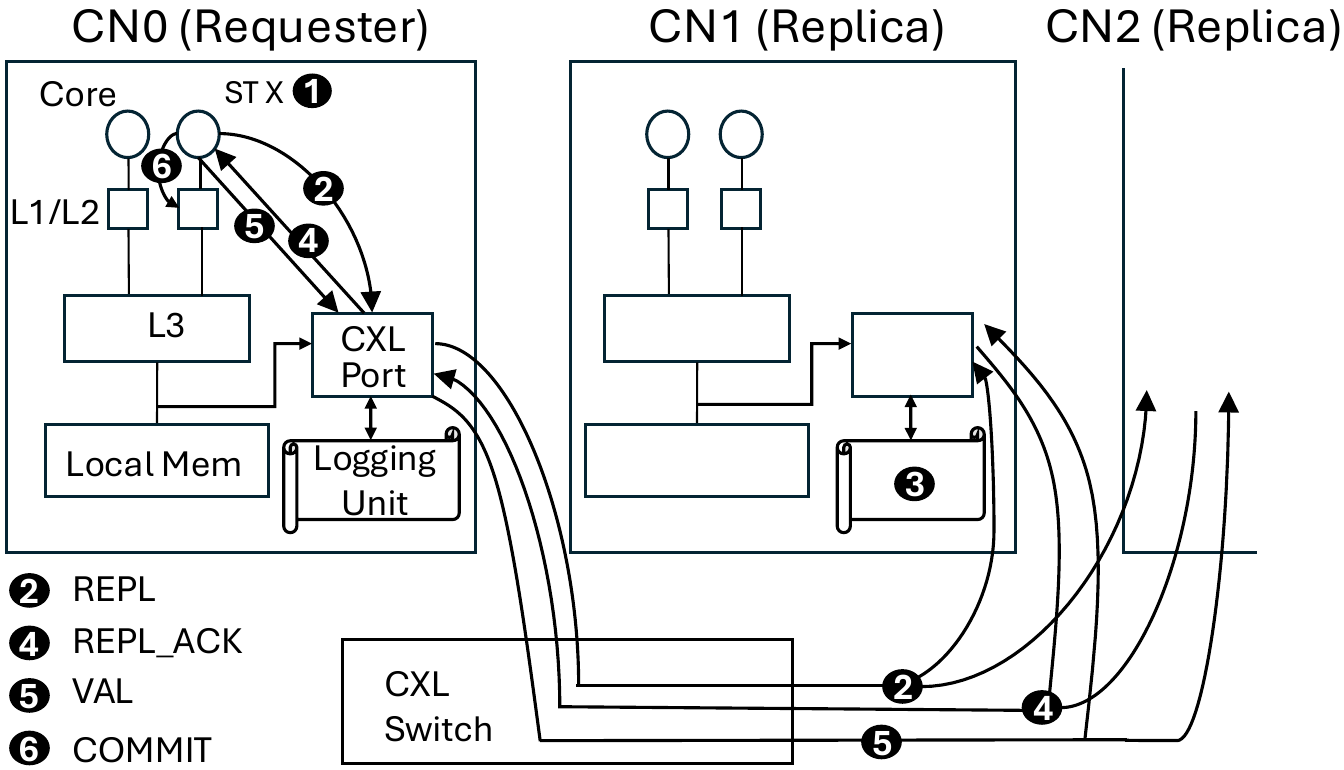} 
    \end{subfigure}
    \caption{\sysname extensions to the transaction of a write 
    to an address in CXL
    memory, assuming two replicas.}
    \label{fig:design:overview}
\end{figure}

Figure~\ref{fig:design:overview} shows the messages added to a write ST X \circleint{1} from a core in CN0 to  a physical address   in 
CXL memory when the system uses two 
replicas. In addition to the regular coherence protocol transaction for the write itself and before the write can commit,
the protocol sends two {\em Replication (REPL)} messages \circleint{2} with the update
to the two replica nodes. The latter save the 
update in the log of their Logging Unit \circleint{3} and respond with a {\em REPL\_ACK} message \circleint{4} to the originating core.
The originating core waits until two events happen: it has 
received all the REPL\_ACKs and the write coherence transaction has completed.
At this point, the protocol: 1) sends  two {\em Validation (VAL)} messages to the replicas \circleint{5} to notify them that replication was successful, and 2)  
commits the write \circleint{6}.
When a log receives the VAL, it marks
the log entry as validated.

The number of replicas per update ($N_{r}$) is a design parameter. 
It is a small value (e.g., 1–4) that depends on the machine size,
trading off 
fault-tolerance and overhead~\cite{minos-hpca24, zookeeper-atc10, dynamo-sosp07, hermes-asplos20}.
To determine which CNs save the replicas for a given cache-line address, the hardware uses a hash 
function on the line address. Hence, all the updates to a given line address are logged in
the same set of $N_{r}$ nodes. It is possible that the writer of a line is in one
of these $N_{r}$ nodes for the line.

Loads are unaffected by \sysnamex. The same stands for writes 
to  physical addresses  in 
the local memories of the CNs. 
 
\vspace{1mm}
\noindent {\bf 2. Error Detection.}
CXL's Link Layer detects errors in devices, switches, and links. 
\sysname uses CXL's Link Layer for error detection and extends it minimally to detect CN failures. 
Specifically, each CXL switch includes an extra Viral\_Status bit for each of the CNs that it is connected to. Additionally,
a switch never responds on behalf of a failed CN, since \sysnamex's goal is correct execution---not just
isolation. 
Section~\ref{sec:fail:detect} describes error detection.

\vspace{1mm}
\noindent {\bf 3. Error Recovery.}
After a  CN error is detected, one of the live cores is informed with a CXL MSI message~\cite{cxlspec} to act  as a Configuration Manager.
Such core brings the workload execution 
to a halt and initiates a distributed software algorithm that disables the faulty CN and recovers the application state to the correct state immediately before the fault. 
Sections~\ref{sec:fail:recov-overview} and \ref{sec::diktamo::dirCntrRecov} describe the recovery.

\subsection{Failure Model}\label{subsec:failure-model}

The focus of this paper is on CN failure, leaving link, switch, and device failure handling to the support already present  
in the CXL specification. 
For CN failures, we assume the standard \textit{fail-stop} model: a faulty CN halts and ceases participation in the protocol. 
This model is widely used in practical crash-fault-tolerant (CFT) systems such as \hbox{RAMCloud~\cite{ramcloud-sosp11}, FaRM~\cite{farm-nsdi14}, and Hermes~\cite{hermes-asplos20}}, as it captures the dominant failure modes in modern data centers (e.g., power loss, kernel panic, or hardware crash).

Byzantine or malicious behavior (e.g., arbitrary message fabrication or equivocation) is out of scope. This assumption is widely consistent with prior work in crash-fault-tolerant replication~\cite{ramcloud-sosp11, farm-nsdi14, hermes-asplos20}, where the dominant concern is reliability against hardware or software crashes rather than adversarial behavior.  We also consider silent data corruption errors out of scope. 

As in prior work~\cite{polardb-cxl, apta-dsn23}, we assume that 
MNs do not suffer faults, and that any data that is written to them is safe.
MNs are bigger, more expensive servers that have stronger RAS support, possibly based on
techniques such as
replication, persistent storage, or the use of a more expensive technologies. How such resiliency is 
accomplished
is not relevant to our work; we simply assume that the MN memory is 
safe at no additional latency cost. 

The  goal of \sysname is not to attain the highest cluster availability. 
Rather, it is to recover the program state
that existed in caches, directory, and memory right before the CN failure. No committed store can be lost as long as at least one replica remains online.

%% file: Sections/diktamo-design.tex
\section{Augmenting Ordinary Execution}
\label{sec:design}

This section describes the  \sysname extensions added to the ordinary execution: the message types, the Logging Unit, the logical timestamps, three variations of the  \sysname replication protocol, and the log processing.

\subsection{\sysname Messages}\label{sec:diktamo:components}

\begin{figure}[t]
    \centering
    \begin{subfigure}[b]{0.6\linewidth}
        \centering
        \includegraphics[width=0.95\linewidth]{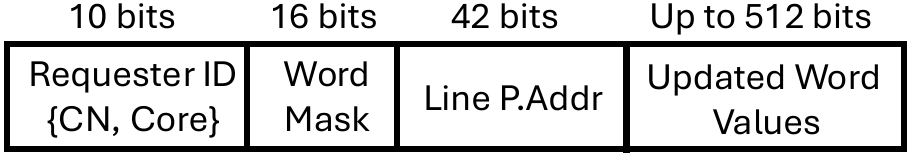} 
        \caption{REPL message} 
        \label{fig:design:replmessage}
    \end{subfigure}
    \begin{subfigure}[b]{0.4\linewidth}
        \centering
        \includegraphics[width=0.99\linewidth]{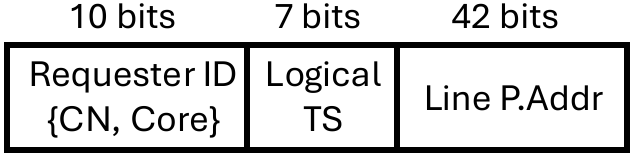} 
        \caption{VAL message} 
        \label{fig:design:valmessage}
    \end{subfigure}
    
    \caption{\sysnamex's \texttt{REPL} and \texttt{VAL} messages.} 
    \label{fig:design:messages}
\end{figure}

\sysname introduces three new protocol messages:  \emph{Replication} (\texttt{REPL}),  \emph{Replication Acknowledgment} (\texttt{REPL\_ACK}),  and \emph{Validation} (\texttt{VAL}). 
The \texttt{REPL} message (Figure \ref{fig:design:replmessage}) is sent by a core in a requester CN  to multiple  replica CNs to replicate a
remote store (or a set coalesced stores to the same cache line). It carries: a) the \emph{Requester ID}, which includes the CN and core IDs, b) a bit mask that shows which words within the line are updated by the store(s) (in this paper, we assume word granularity, but the design can be modified to use finer granularity), c) the physical address of the line, and d) the updated values of the words. The last field is of varying size, depending on the number of updates that are  coalesced. The maximum size is a cache line size. 

The \texttt{REPL\_ACK} message is sent by the \emph{Logging Unit} of each replica CN back to the requester core. It is sent after the unit has applied the \texttt{REPL} update to its  hardware log.  Upon receiving all \texttt{REPL\_ACK}s, the requester core knows that replication has reached all replicas. 

The \texttt{VAL} message (Figure \ref{fig:design:valmessage})  is sent by the requester core to all replica CNs after it receives all their \texttt{REPL\_ACK}s. \texttt{VAL} indicates that the write is committed and so the Logging Unit should record the update as such.
\texttt{VAL} carries the Requester ID  and the target cache line address. It also carries a logical timestamp (TS),
which is used to order the updates as they are logged in the Logging Unit (Section~\ref{sec:design:ts}).

\subsection{Logging Unit} 
\label{sec:design:offloadingunit}

Each CN has a hardware Logging Unit that logs the data of incoming \texttt{REPL} messages, compresses the data  and then, at regular intervals,
sends it to the MNs, where it is saved.
The unit is linked to the CN's CXL port (Figure~\ref{fig:design:overview}). 

\begin{figure}[t]
    \vspace{-2mm}
    \centering
    \includegraphics[width=0.6\linewidth]{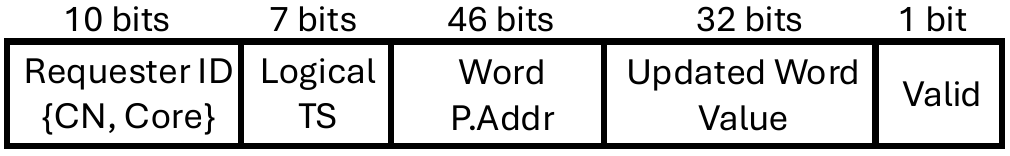} 
    \caption{\sysname   log entry in the SRAM Log Buffer.} 
    \label{fig:design:logentry}
\end{figure}

The Logging Unit's log is made of DRAM memory for density. To improve access speed, however, incoming  \texttt{REPL} and
\texttt{VAL} messages update a small SRAM memory called the {\em SRAM Log Buffer}.
A \texttt{REPL} allocates an entry in the SRAM Log Buffer.
Later, the corresponding \texttt{VAL} sets the Valid bit of the entry, and
the  entry 
becomes eligible to be moved to the DRAM log in the background. 
In there, it will be compressed and eventually
sent to an MN.
Non-VALed entries correspond to writes that have not yet committed. These entries cannot be sent to the DRAM log. 
Persistency of the log is not required, since
\sysnamex{} employs a replication factor of $N_{r}$. This allows the system to tolerate up to $N_{r}$-1    CN failures 
without losing state. 

Figure~\ref{fig:design:logentry} shows the layout of an  entry
in the SRAM Log Buffer. Each entry includes the data of a single 
remote store instruction. If a received \texttt{REPL}  message carries multiple stores to the same line, then the Logging Unit splits the message into an equal number of log entries. Each entry has the Requester ID, a logical timestamp, the  physical address and  value of the updated word, and a valid bit. 
The latter is set only when the unit receives the \texttt{VAL} message. 

\subsection{Logical Timestamps} 
\label{sec:design:ts}

The \sysname protocol decouples the allocation of a log entry in the SRAM Log Buffer from the saving of the entry into the DRAM log. The former is performed on reception of a \texttt{REPL},
while the latter can only happen after the reception of the \texttt{VAL}. Now, consider the case where the cores in a given CN perform  two updates to the same 
memory location. The \texttt{VALs} are issued in the order of the updates---since they are issued at the point
when an update commits. However, the CXL fabric could potentially reorder the two \texttt{VAL}s
and they could reach the 
SRAM Log Buffer out of order. 
To ensure that the log knows the order of the \texttt{VAL}s, 
we add a 7-bit logical timestamp (TS)
to the \texttt{VAL} messages.
TS is incremented every time a \texttt{VAL} is sent.
In the logging CN, the SRAM Log Buffer saves the \texttt{VAL}'s TS in the log entry.
However, it only pushes the entries from that word into the DRAM log in timestamp order. This ensures that  the recovery will not 
apply the updates in the wrong order.

\begin{figure}[t]
    \centering
    \includegraphics[width=0.85\linewidth]{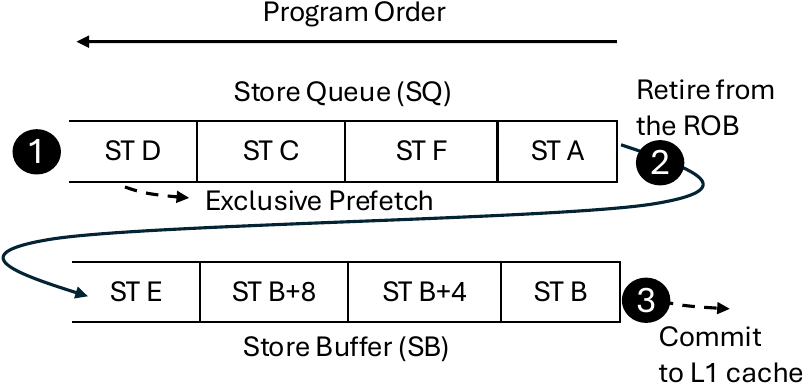} 
    \caption{Multiple stores in an out-of-order core with TSO.} 
    \label{fig:design:microarch}
\end{figure}

\begin{figure*}[t]
    \centering
    \begin{subfigure}[b]{0.31\linewidth}
        \centering
        \includegraphics[width=0.92\linewidth]{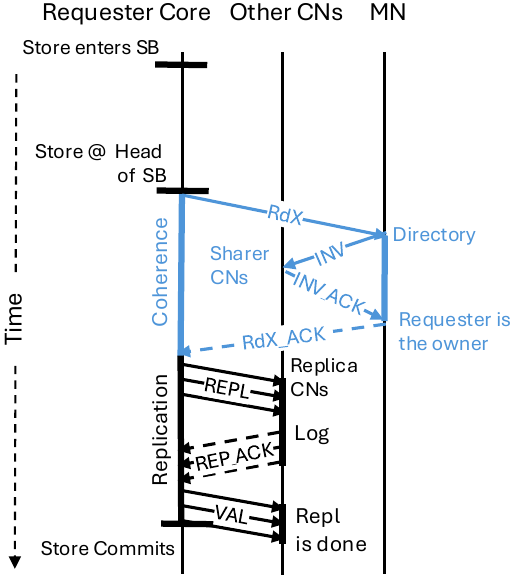} 
       \caption{\sysnamebaselinex}
       \label{fig:design:diktamoB}
    \end{subfigure}
    \hfill
    \begin{subfigure}[b]{0.31\linewidth}
        \centering
        \includegraphics[width=0.89\linewidth]{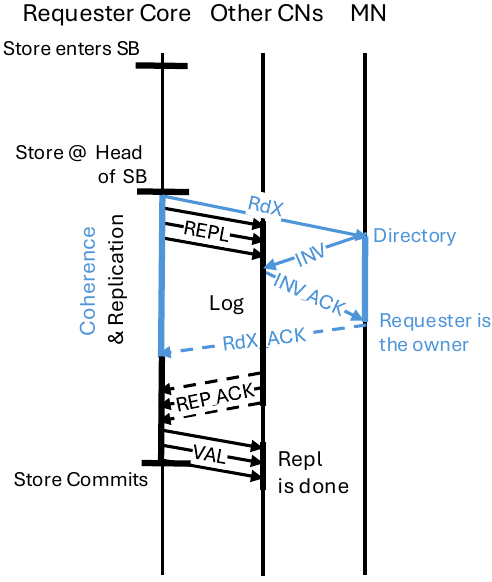} 
        \caption{\sysparallelx}
        \label{fig:design:diktamoPa}
    \end{subfigure}
    \hfill
    \begin{subfigure}[b]{0.31\linewidth}
        \centering
       \includegraphics[width=0.89\linewidth]{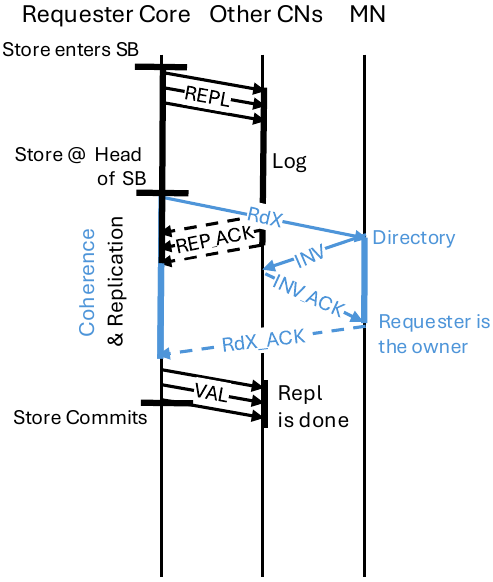}
       \caption{\sysproactivex}
       \label{fig:design:diktamoPr}
    \end{subfigure}
    \caption{Timeline of the Coherence and Replication transactions of a store in the three variations of the \sysname protocol. This timeline assumes that an early 
    exclusive prefetch of the line targeted by the store has been ineffective 
    because the line has since been invalidated, downgraded, or evicted from L1. In cases where the exclusive
    prefetch has been issued but not completed by the time the
    store reaches the head of the SB, the Coherence transaction
    will only observe partial latency. If the exclusive prefetch has completed and the data is in the L1, the Coherence
    transaction will take no time.}
    \vspace{-4mm}
    \label{fig:design:diktamo}
\end{figure*}

\subsection{Three Variations of \sysnamex's Protocol}\label{sec:diktamo:replication}

In general, when a store is ready to commit, it issues a message that  invalidates
all other cached copies of the line and brings 
the line 
to the local cache. 
We call this operation the {\em Coherence} transaction.
\sysname augments the coherence transaction with a {\em Replication} transaction, which involves
sending multiple  \texttt{REPL}s, receiving multiple \texttt{REPL\_ACK}s, and sending multiple
\texttt{VAL}s. Only when both the Coherence and the Replication transactions for a store complete
can the store commit---i.e., merge the update into the local cache. In our design, we propose three variations of 
\sysname that overlap the Coherence and the Replication transactions for a store differently.
Before we present them, we describe the life of a store instruction.

\vspace{1mm}
\noindent {\bf 1. The life of a Store Instruction.}
Figure~\ref{fig:design:microarch} shows multiple store instructions in an out-of-order core with TSO. For ease of explanation, we use two
logical structures: the store queue (SQ) and the store buffer (SB). When a store enters the reorder buffer (ROB), it also enters the SQ. When its target physical address is resolved, an exclusive prefetch is sent to memory to bring the target cache line in Exclusive state to the L1 cache of the core~\circleint{1}. When a store reaches the head of the ROB (and hence of the SQ), it retires into the SB~\circleint{2}. With the 
TSO  
memory consistency model~\cite{x86TSO-cacm10,sparcTSO-94},
stores  drain from the SB in order, without overlap: a store is merged with the 
memory system
only when it reaches the head of the SB~\circleint{3}. We will call this point  the  store commit. 

A common optimization  is \emph{store coalescing}. Consecutive stores in the SB that target different words of the same cache line and are not interleaved by a store to a different line are coalesced and committed to memory in a single transaction. For instance, 
in Figure~\ref{fig:design:microarch}, if stores ST B, ST B+4, and ST B+8 write
different addresses of the same line, they 
will be coalesced and merged with memory together. In such cases, \sysname sends
a single \texttt{REPL} for all the coalesced stores. In the \texttt{REPL} (Figure~\ref{fig:design:messages}a), the Word Mask
sets the corresponding bits and the Values field includes all the updated word values.
This reduces the network traffic of replication. 
All three \sysname designs support coalescing.

If the exclusive prefetch for a store 
successfully brings the target line into the L1 cache
and the line  has not been invalidated, downgraded or evicted by the time the store reaches the head of the 
SB, then what we call the Coherence transaction takes no time. 
However, for illustration
purposes, in the following discussion, we will assume that the line has 
been invalidated, downgraded, or evicted  and, when 
the store reaches the head of the SB, the Coherence transaction 
experiencies the full latency. In cases where the  exclusive prefetch
has been issued but not completed by the time the store reaches the 
head of the SB, the Coherence transaction will only observe partial latency.

\vspace{1mm}
\noindent {\bf 2. First Design: \sysbaselinex.}
In this design, the Replication transaction of a store starts only after the Coherence transaction has completed. When the Replication transaction completes, the store  can commit. 

Figure~\ref{fig:design:diktamoB} shows the timeline of a store from the time it enters the
SB in the requester core until it commits. It includes the operations in the requester core, in other CNs, and in the MN that contains the directory. It shows the Coherence transaction in blue and the Replication one in black.
We see that, when the store reaches the head of the SB, the Coherence transaction starts.
We show the general case where the line is currently cached in  Shared state by  other nodes.
A request for ownership (RdX) is sent, which reaches the directory in an MN. The latter sends invalidations
to all sharer CNs, collects their ACK responses, marks the requester as the owner, and responds 
with the cache line (RdX\_ACK).

The Replication transaction starts now. The CN with the requester core   sends
\texttt{REPL} messages to replica CNs (three in the example).
Each replica CN logs the update and replies with a \texttt{REPL\_ACK} to the requester. When the requester receives all the \texttt{REPL\_ACK}s, it sends the \texttt{VAL}s   
to all the replicas and commits the store---i.e., it drains the store from the SB and 
updates the cache line in L1. 

\vspace{1mm}
\noindent {\bf 3. Second Design: \sysparallelx.}\label{sec:diktamo:parallel}
In this design, the Replication transaction of a store starts at the same time as the Coherence transaction when the store reaches the head of the SB. Both transactions overlap in time. Figure~\ref{fig:design:diktamoPa} shows the timeline of a store. Note that \sysparallel  can only send the 
\texttt{VAL}s (and immediately commit the store) when two conditions are met: 1) the originating core 
has received all the \texttt{REPL\_ACK}s and 2) the Coherence transaction has completed.
The advantage of this design
is that, by overlapping the two transactions, the store  commits sooner.

\vspace{1mm}
\noindent {\bf 4. Third Design: \sysproactivex.}
In both \sysbaseline and \sysparallelx{}, replication is initiated no earlier than when a store reaches the head of the SB. As a result, replication is {\em serialized} across stores. However, all the necessary information for the replication is 
known at least as soon as the store retires and enters the SB (\circleint{2} in Figure~\ref{fig:design:microarch}). To send the \texttt{REPL}s, we do not need to wait for the many cycles that a store may wait in the SB until it reaches the SB head~\cite{systematic-asplos25}. 
 
To avoid waiting in the SB, we introduce  \sysproactivex, where the \texttt{REPL}s
for a store are sent out as soon as the store retires and  enters the SB.
Figure~\ref{fig:design:diktamoPr} shows the   store timeline. 
As before, 
\sysproactive  can only send the 
\texttt{VAL}s (and immediately commit) when both the \texttt{REPL\_ACK}s are received 
and  the Coherence transaction has completed.

Sending the \texttt{REPL}s earlier (e.g., when the exclusive prefetch is issued) is 
undesirable because the store is speculative. As a result, the store could be squashed
due to a misspeculation, which would lead to two issues: (a) unnecessary network
traffic, and (b) nonVAL-ed entries in logging units that would require a dedicated 
mechanism to remove. Unlike an exclusive prefetch, a squashed Replication transaction
does not have a  useful side effect.
 
\begin{figure}[t]
    \vspace{-1mm}
    \centering
    \includegraphics[width=\linewidth]{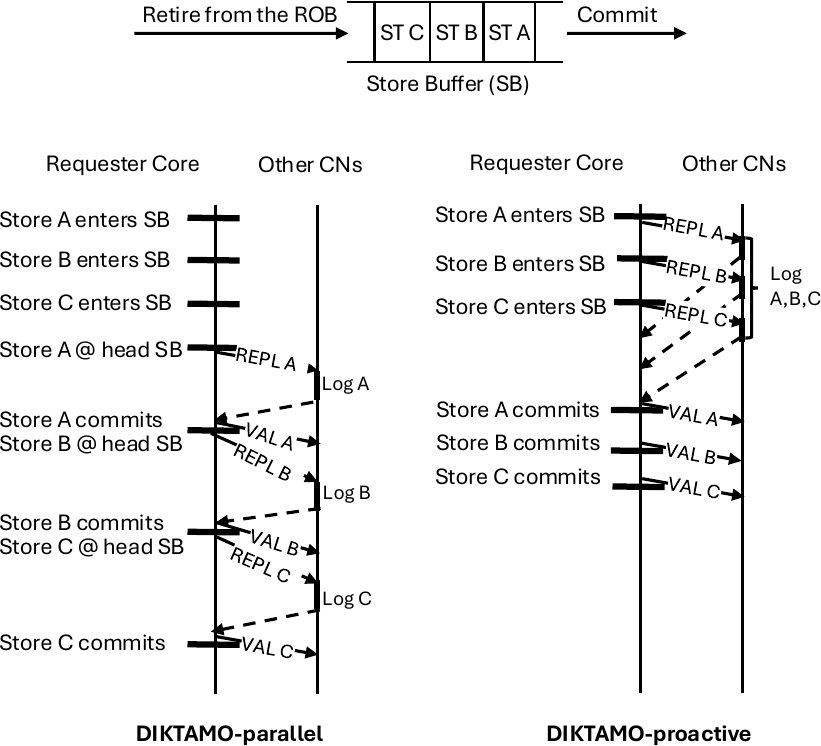} 
    \caption{\sysproactive overlaps the \texttt{REPL}--\texttt{REPL\_ACK} cycle for
    multiple retired stores waiting in the SB.}
    \label{fig:design:multiplestores}
    \vspace{-1mm}
\end{figure}

\sysproactive has two  performance advantages. First, it hides the latency of replication behind the existing SB queuing delays. 
{\em Second and crucially, it  overlaps the \texttt{REPL}--\texttt{REPL\_ACK} cycles of 
multiple retired stores waiting in the SB, delivering faster execution without violating TSO}. 
To see how, Figure~\ref{fig:design:multiplestores} shows the case of stores ST A, ST B, and ST C
waiting in the SB. The leftmost chart shows the timeline for \sysparallelx. 
Each store  waits until it reaches the head of the SB and then performs the Replication transaction in sequence. 
In contrast, the rightmost chart shows the timeline for \sysproactivex.
It issues \texttt{REPL}s for consecutive stores in order, but allows their 
\texttt{REPL}--\texttt{REPL\_ACK} cycles to fully overlap.
\sysproactive can support as many outstanding replication requests as the size of the SB, without
violating TSO.  

\vspace{1mm}
\noindent {\bf 5. \sysproactive and Store Coalescing.}
\sysbaseline 
and \sysparallel trivially support store coalescing: 
as the two (or more) oldest stores in the SB are coalesced, only a single
\texttt{REPL} message set is sent for all of them.
However, \sysproactive requires some subtle redesign to support coalescing. This is because
individual stores  send their own \texttt{REPL} sets as soon as they retire
and are placed in
the SB, 
while, later on,  coalescing may potentially 
merge multiple of them  when they are draining from the SB. The result
will be  a single multi-store commit operation and, hence, a single \texttt{VAL}.
This would be confusing and require supporting special cases.

Consequently, to support coalescing, we slightly modify \sysproactive as follows.
When a store $B$ is deposited into the SB, it never issues its \texttt{REPL} set. If it finds that
it is immediately preceded by
a store $A$ to another word of the same line, it is coalesced with $A$; 
no other action is taken. Later, when the next store $C$ is deposited into the SB, if $C$ cannot be
coalesced with $B$, $B$ issues its \texttt{REPL} set (which carries $B$ and all the earlier 
stores coalesced with $B$).
Finally, if a store $B$ reaches the head of the SB without having sent 
its \texttt{REPL} set, it
sends it at this point (carrying $B$ and all the stores coalesced with $B$). This design 
supports coalescing, while initiating only a single \texttt{REPL}s-\texttt{REPL\_ACK}s-\texttt{VAL}s transaction for each commit.

\subsection{\sysnamex's Log Processing}
\label{sec:diktamo:logprocessing}

The Logging Units of all the CNs continuously compress the logged data and,
at periodic intervals, they all 
save  the compressed data 
to the MNs, where the data is 
safe from faults. Given that there are $N_r$ replicas, groups of $N_r$ nodes   have the same data
in their logs. We call these groups of CNs {\em Replica Groups}. To reduce   overhead, 
the Logging Units in a Replica Group divide their work. Specifically, in a Replica Group,
each of the Logging Units is in charge of saving only log entries within a range of physical 
addresses.

The process followed by each Logging Unit is as follows. The hardware reads the DRAM 
log in sequence, from older to newer entries, extracting the log entries that this particular
Logging
Unit is in charge of saving (based on
the physical address).
As these log entries are being extracted, the hardware compresses them to save memory space. For compression, we use gzip~\cite{gzip-96} with compression level 9, 
which attains an average compression factor of 5.8x in our experiments.
At regular intervals, the Logging Units synchronize through the controllers
in the MNs,
save their compressed logs in 64-byte messages to the MNs,  synchronize 
again, and clear the logs.

The size of the log and the interval between flushes are
set so that logs typically do not overflow. In rare
cases, it is possible that a log in a node reaches a high watermark prior to 
reaching the time to flush the log.
In this case, the Logging Unit in the node triggers an Early Flush interrupt to  the Logging Units of all the other nodes.
Then, all the Logging Units synchronize and
perform the flushing operation.

%% file: Sections/diktamo-failures.tex
\section{\sysname and Failures}\label{sec:failure-handling}

This section describes how \sysname detects a CN failure and outlines the \sysname recovery.

\subsection{Failure Detection Mechanism}
\label{sec:fail:detect}

The Link Layer in CXL detects both transient and permanent errors in devices, switches, and 
links~\cite{cxlspec}. \sysname reuses this support for all the  messages it adds to the protocol. 
However, since the Link Layer does not 
support CN failures (Section~\ref{sec:background:reliability}),
\sysname extends it as follows. First, the CXL specification indicates
that CXL switches contain Viral\_Status bits for various ports to indicate failure conditions in those
ports~\cite{cxlspec}.
\sysname extends the switch design to have one Viral\_Status bit for each CN that is connected to
the switch.
When the switch detects that a given CN is unresponsive, it sets the 
corresponding Viral\_Status bit.  

Additionally, the CXL specification indicates that, when a
message is sent to a failed device, 
the device or nearest switch responds with a message with 
{\em poisoned} or {\em viral} data~\cite{cxlspec}. 
This is done so that the failed device is isolated.
However, in an environment where multiple CNs share data, one cannot 
afford responding 
with incorrect data, as it could pollute the state of the 
running application.
Consequently, in \sysnamex, the switch connected to a
failed CN will not respond at all to requests to the failed node (and the latter will not 
respond either
because of the failure model we assume). 

Instead, as soon as a switch detects a failed CN, it sends a Message Signaled Interrupt (MSI)
to a core in  one of the live CNs to trigger a recovery. An MSI is the standard
mechanism  used by CXL devices to signal interrupts to a host CPU~\cite{cxlspec}.
They function identically to how they are used in PCI Express (PCIe). The receiver
core, which we call the {\em Configuration Manager} (CM) for this fault, will trigger  a coordinated pause of all the CNs,
and then initate a software-driven 
recovery.

\subsection{Recovery Protocol Overview}
\label{sec:fail:recov-overview}

The objective of the software-driven
coordinated recovery is to reconstruct the consistent state that the application had immediately
before the fault, and 
restart the application on the remaining functioning CNs. 
The process involves accessing the  directory state in the MNs and identifying the memory lines
that the directory records as being present in the caches of the failed CN, either in  state \texttt{Shared} or   \texttt{Modified}. For the former, the directory entry must be 
updated by removing the failed CN as a sharer. 
For the latter, the logs in the MNs and in the 
Logging Units must be 
searched to find  the  latest update before the crash. Then, such update must be applied to
memory, and the directory entry must be changed to indicate that there are no sharers.
These operations are performed by recovery software. Our main concern is to 
guarantee correctness and minimize complexity.

\begin{figure}[t]
    \centering
    \includegraphics[width=.90\linewidth]{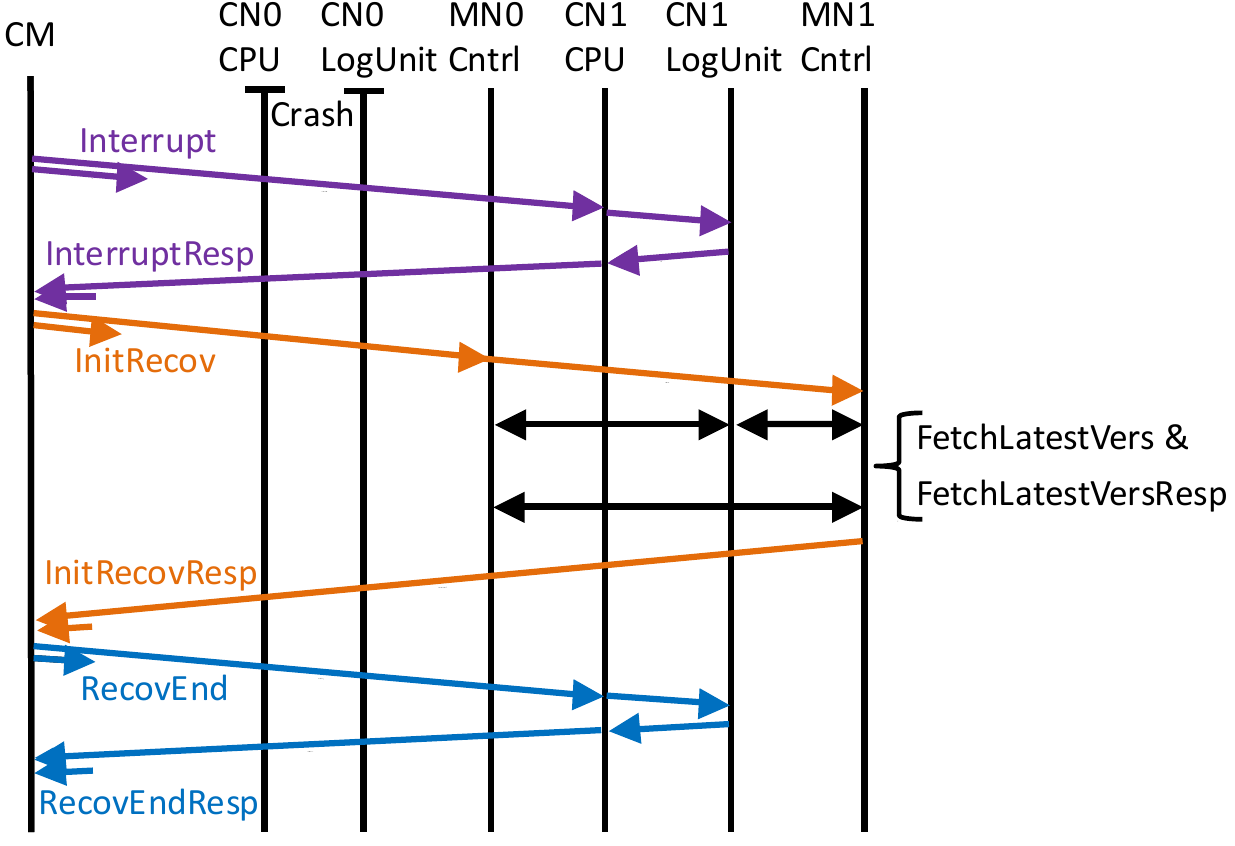}
    \caption{Message exchange during recovery in \sysnamex.}
    \label{fig:recovery-msg}
\end{figure}

Figure~\ref{fig:recovery-msg} shows the messages exchanged by the nodes during recovery. 
Table~\ref{tab:recov:recovery-messages} describes the messages. 
These messages can leverage the mailbox interfaces in the CXL RAS API (v0.8)~\cite{ocp-rasapi23}, which enable host–device communication for reliability and error-handling events.
The recovery process begins with the CM broadcasting an \texttt{Interrupt} to all live CNs  in the   cluster. 
These messages are directed to all the CPU cores in the CNs  and to the per-CN Logging Units.
Upon receiving the interrupt, cores and Logging Units
complete all outstanding
requests/operations, and pause.
Then, they respond with an \texttt{InterruptResp}.
Once all \texttt{InterruptResp}s have been received, the CM issues an \texttt{InitRecov} to all MNs, prompting software  
handlers in the directory controllers of the MNs to initiate recovery. The actions followed by 
these handlers   set the directories and memories to a consistent state.
They are described in Section~\ref{sec::diktamo::dirCntrRecov}.

\begin{table}[t]
    \footnotesize
    \centering
    \caption{Recovery protocol messages.}
    \begin{tabular}{|l|p{4.8cm}|}
    \hline
    \textbf{Message} & \textbf{Definition / Usage} \\
    \hline
    \texttt{Interrupt} & Sent by the CM to all live CNs and their Logging Units to complete outstanding requests and other operations in the cores and in the Logging Units. \\
    \hline
    \texttt{InterruptResp} & Sent by CNs to confirm completion of outstanding requests and operations. \\
    \hline
    \texttt{InitRecov} & Issued by the CM to all MNs to initiate directory-level recovery actions by software handlers. \\
    \hline
    \texttt{FetchLatestVers} & Request from the directory controllers to the Logging Units and MNs for the latest logged updates to the lines that were owned by the failed CN's caches. \\
    \hline
    \texttt{FetchLatestVersResp} & Response with the latest logged updates to   the lines that were owned by the failed CN's caches. \\
    \hline
    \texttt{InitRecovResp} & Issued by  handlers in directory controllers to acknowledge completion of directory-level recovery actions. \\
    \hline
    \texttt{RecovEnd} & Broadcast by the CM to all live CNs and their Logging Units to signal that recovery is complete and they can resume execution. \\
    \hline
    \texttt{RecovEndResp} & Sent by CNs to acknowledge completion of recovery and execution resumption. \\
    \hline
    \end{tabular}
    \label{tab:recov:recovery-messages}
\end{table}

Once the directory and memory in an MN have been brought
to a consistent state, the handler in the directory controller sends an 
\textit{InitRecovResp}   to the CM. 
Upon the reception of \textit{InitRecovResp} from all live directory controllers, the CM broadcasts a \texttt{RecovEnd} to all CNs and Logging Units, signaling the completion of the recovery protocol. Each node responds with a \texttt{RecovEndResp} and resumes program execution.

\begin{algorithm}[t]
    \footnotesize
    \caption{Recovery of the Directory and Memory State by a Software Handler at the Directory Controller}
    \KwIn{Set of dir entries marked as owned or shared by  failed CN}
    \KwOut{Corrected directory and memory state}

    For all dir entries where the failed CN was a sharer: \\
    {\hspace*{1em}}Remove the failed CN as sharer\;
    
    For all dir entries where the failed CN was the owner: \\
    {\hspace*{1em}}Collect word addresses of corresponding lines in \texttt{Owned} list\;
    Send \texttt{FetchLatestVers} with \texttt{Owned} list to relevant LUs\;
    Collect all \texttt{FetchLatestVersResp}\;
    Identify LatestVersions in all \texttt{FetchLatestVersResp}\;
    Remove addresses of identified LatestVersions from \texttt{Owned} list\;
    If \texttt{Owned} list not empty:\\ 
    {\hspace*{1em}}Send \texttt{FetchLatestVers} with \texttt{Owned} list to relevant MNs\;
    {\hspace*{1em}}Collect all \texttt{FetchLatestVersResp}\;
    {\hspace*{1em}}Identify LatestVersions in all \texttt{FetchLatestVersResp}\;
    Write identified LatestVersions from logs in MNs and LUs to mem\; 
    Mark their directory entries as not owned\;
    \label{sec:fail:algo-directory}
\end{algorithm}

\subsection{Recovery  at the Directory Controller}
\label{sec::diktamo::dirCntrRecov}

Each directory controller runs  a software  handler that repairs the  
directory and memory state in the local MN. The high-level algorithm is shown in   Algorithm~\hbox{\ref{sec:fail:algo-directory}}. The handler first traverses
the local directory and performs two operations. For the
directory entries where the failed CN was a sharer, it
removes the failed CN as a sharer. For the directory entries where the failed CN was the owner,
it collects the addresses of all the words in the corresponding lines 
in a list called \texttt{Owned}.

The handler then proceeds to collect  the latest versions of all the words in \texttt{Owned}.
This is done in two steps. First, the handler sends a
\texttt{FetchLatestVers} message with the \texttt{Owned} list to all the
 Logging Units that could have updates to such words. 
In each Logging Unit, a handler searches for them and 
returns a \texttt{FetchLatestVersResp}
message. The message contains, for each requested word address, the 
latest logged update  to it.
When the directory receives the \texttt{FetchLatestVersResp} 
from all replicas, it 
compares them. For each address  in   the  \texttt{Owned} list, two cases are possible.
One case is when all the replica logs have an update (which is the same in all
replicas).
If so, this is the latest version. 
A second case is when none of the replica logs have an update. 
In this case, the version is in the MN logs.

In the second step, the directory controller removes from the 
\texttt{Owned} list all the addresses for which it has collected the 
latest versions. It then sends a
\texttt{FetchLatestVers} message with the \texttt{Owned} list to all the
MNs whose log could have the missing  versions. In each MN, a software handler searches
the log and returns the latest version  of  the addresses in a \texttt{FetchLatestVersResp}. When the directory controller handler receives the \texttt{FetchLatestVersResp} from all replicas, it finds the latest versions 
of the
remaining words. Then, the handler collects all the latest versions
found in the two steps, writes them to  memory in the local MN, and updates the 
corresponding local directory entries so the lines 
are not owned  by any CN.

%% file: Sections/methodology.tex
\section{Methodology}
\label{sec:methodology}

\noindent{\bf Architecture and Simulation Infrastructure.}
We simulate a distributed system with 16 compute nodes (CNs) of four cores each, 
and 16 memory nodes (MNs), interconnected via high-bandwidth CXL links.
The architecture parameters are shown in Table~\ref{tab:sst-config}. There is
a single CXL switch connecting all CNs to all MNs. 
Cores are out-of-order with private L1/L2 caches. All cores in a CN share an 
inclusive L3 cache. Logging units 
operate at 500~MHz.
Each Logging Unit has a 4KB SRAM Log Buffer and a 18 MB DRAM log. 
We model CXL using parameters from the literature~\cite{pond-asplos23, micron-cxl, marvell2026cxl}. We assume that the MNs 
have features that make them
fault tolerant. We follow  prior work~\cite{polardb-cxl, apta-dsn23} in this assumption. 

\begin{table}[t]
    \caption{Architecture parameters. RT means round trip.}
    \label{tab:sst-config}
    \centering
    \resizebox{.99\linewidth}{!}{
        \begin{tabular}{||l|l||}
            \hline\hline
            \multicolumn{2}{||c||}{System Configuration} \\
            \hline\hline
            Compute Nodes (CN); Memory Nodes (MN) & 16 CNs; 16 MNs \\
            CPU Frequency; Logging Unit Frequency           & 2.4~GHz; 500~MHz \\
            Cores per CN              & 4 Out-of-Order cores\\
            Load Queue Size; Store Queue Size               & 128; 72 \\
            \hline\hline
            \multicolumn{2}{||c||}{Cache Hierarchy \& Local Memory (per CN)} \\
            \hline\hline
            L1 Cache (Size; Associativity; RT Latency)       & 48~KiB; 12-way; 5 cycles \\
            L2 Cache (Size; Associativity; RT Latency)       & 512~KiB; 8-way; 13 cycles \\
            L3 Cache (Total Size; Associativity; RT Latency) & 8~MiB; 16-way; 36 cycles \\
            Cache Line Size & 64 bytes\\
            \hline\hline
            \multicolumn{2}{||c||}{CXL Memory and Interconnect} \\
            \hline\hline
            DRAM Access Latency           & 45~ns   \\
            CXL Switch Bandwidth (Aggregate) & 4~TB/s~\cite{marvell2026cxl}\\
            CXL Network RT Latency (includes Dir  Lookup)    & 200~ns~\cite{pond-asplos23} \\
            \hline\hline
            \multicolumn{2}{||c||}{Logging Unit
            } \\
            \hline\hline
            Log Buffer Size  &  SRAM: 4KB (4ns access);  \\
                      &  DRAM: 18MB\\
            Period between Log Dumpings & 2.5 ms \\
            Replication Factor  & $N_{r}$=3  copies  \\
            \hline\hline
        \end{tabular}
    }
    \vspace{-2mm}
\end{table}

We use the Structural Simulation Toolkit (SST)~\cite{sst-sigmetrics11} to model 
the architecture.
Execution traces of the program execution are collected  per core using Pin~\cite{intel-pin}, and fed into the simulator. 
The traces include all 
data accesses, including
synchronizations (lock acquire request, lock release, barrier arrive, and barrier leave). In between every two 
consecutive entries, 
the trace also includes K\_$i$, the number of other types of instructions executed between the two accesses.  In the
simulations, each core advances time by   adding K\_$i$ cycles in between 
memory operations 
(which assumes that non-memory instructions have a CPI equal to 1). 
We ensure that the synchronizations are correctly modeled. Specifically, only one thread can be inside a given critical section at a time (the others can be spinning), and threads spin on a barrier until all other threads 
have arrived. 
We compile the applications with -O3 to eliminate unnecessary local accesses.

\noindent{\bf Applications.}
We use a mix of 
applications from PARSEC~\cite{parsec-pact08} and SPLASH-2~\cite{splash2-isca95, parsec-splash2-iiswc08}, and a custom key-value workload using Yahoo! Cloud Serving Benchmark (YCSB)~\cite{ycsb-socc10, ycsb-code-2014}. 
From PARSEC, we use \emph{bodytrack}, \emph{fluidanimate}, \emph{streamcluster}, and \emph{canneal}, which are a range of animation, computer vision, engineering, and data mining workloads.
From SPLASH-2, we use \emph{raytrace}, \emph{barnes}, \emph{ocean\_ncp}, and \emph{ocean\_cp}, which are 
workloads that stress memory-level parallelism. 
For YCSB, we use
a distributed key-value store backed by a hashtable, inspired by Bigtable~\cite{bigtable-acm08}. 

All applications are multi-threaded and are executed with 64 threads running on all cores and fully utilizing the cluster. For each application, we execute
  6.4B instructions or more. 
For the PARSEC and
SPLASH-2 applications, we skip the initialization and start the evaluation when the application reaches the parallel section or the Region of Interest. We use the Medium size of the applications or larger. We place the shared data  on the CXL memory, and the private data of the threads on the local memory of the CNs

For YCSB, the key value store has 
1.6M records of size 1KB. 
The application utilizes a database organized in an array-like format. 
It performs 6.4 million put or get record operations
to the CXL memory.
It issues 80\% reads and 20\% writes with a uniform record access distribution. 
We place all the key value store data on the CXL memory.

\noindent{\bf Configurations.}
\label{sec:methodology:configurations}
We evaluate five architecture configurations that handle the accesses
to the data allocated in CXL memory (i.e., remote data) differently.

\noindent $\bullet$ \textbf{Write-Back (WB)}: Conventional, non-resilient system where  stores to remote data (i.e., remote stores) are committed to local caches without any replication and without simultaneously saving the update in any MN.
This configuration is not resilient: the failure  of a  CN  results in data loss. It provides a lower-bound execution time.

\noindent $\bullet$ \textbf{Write-Through (WT)}: 
Resilient system where remote stores are write through, updating the memory in a  MN. As soon as a write
updates a MN's DRAM memory, we assume that the update cannot be lost---since the MN is fault tolerant. 

Under the TSO memory model~\cite{x86TSO-cacm10,sparcTSO-94}, multiple write-through updates from a core cannot overlap with each other.

\noindent $\bullet$ \textbf{\sysnamebaselinex}: 
Baseline version of our resilient write-back \sysname protocol, where remote stores are replicated to $N_{r}$ CNs after their coherence transactions complete.
    
\noindent $\bullet$ \textbf{\sysparallelx}: Variant that optimizes \sysnamebaseline by overlapping the issue of the replication and coherence transactions of a store. 

\noindent $\bullet$ \textbf{\sysproactivex}:
Proposed \sysname{} variant, which improves on \sysparallel by allowing the initiation of the replication transactions for remote stores as they enter the SB.
    
All \sysname configurations use the same
scheme for selecting replica CNs: 
a hash function maps the line address to a  Replica Group. $N_r$ is  
set to 3, so that every remote store is replicated in three distinct Logging Units across the cluster. This number is  very conservative (i.e., large) for our small 16-CN machine, given that it is
similar to the number used in prior works evaluating datacenter-scale   distributed systems~\cite{dynamo-sosp07, hadoop-msst10}.
We use this number so we can model the additional traffic that \sysname 
would add in a large machine.

%% file: Sections/evaluation.tex
\section{Evaluation}
\label{sec:eval}

In this section, we first evaluate the performance of \sysnamex, then
characterize \sysnamex, evaluate recovery, and finally perform
sensitivity analyses.

\subsection{Performance}

\begin{figure}[t]
\centering
\vspace{-3mm}
\begin{subfigure}[b]{\linewidth}
    \centering
   \includegraphics[width=\linewidth]{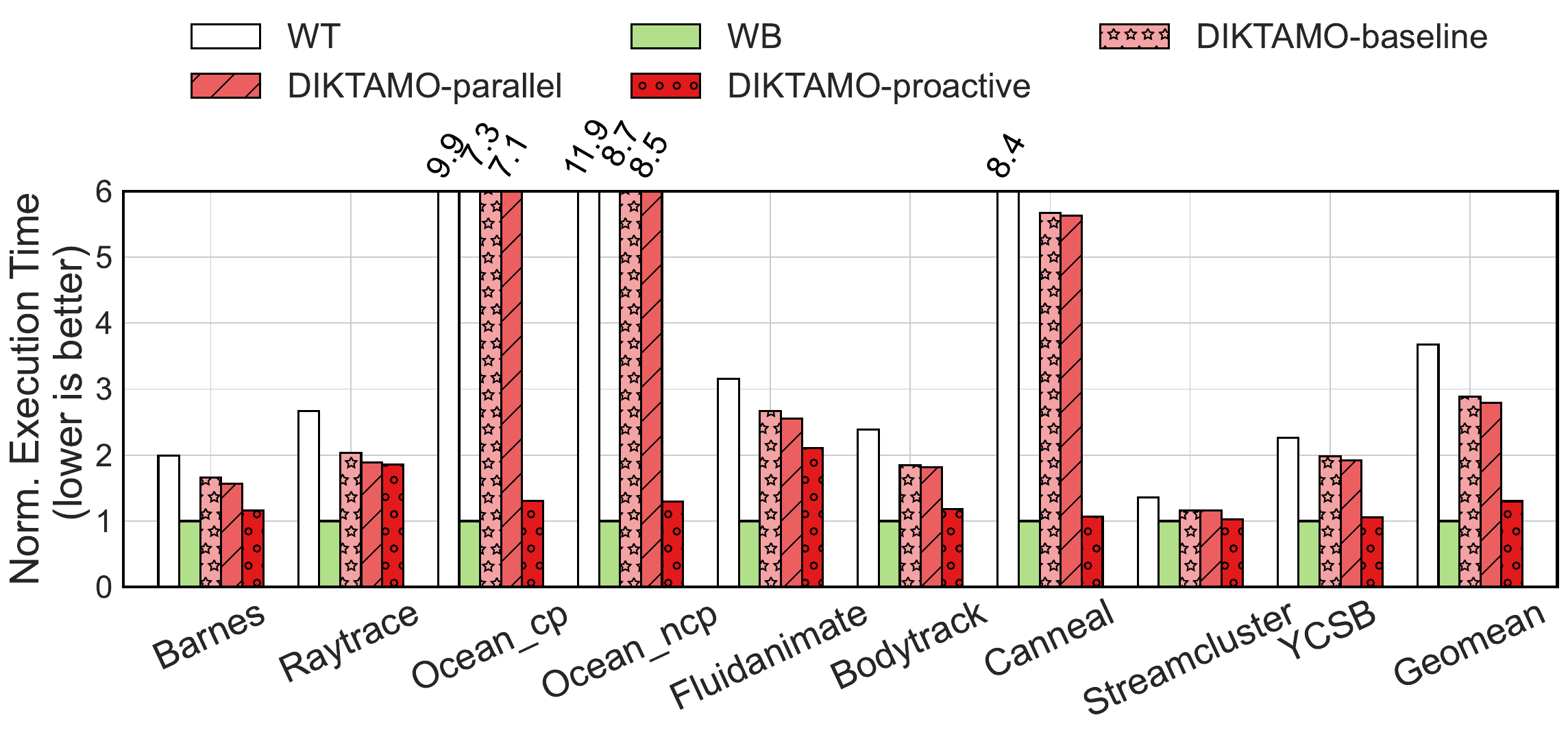}
\end{subfigure}
\caption{Execution time with different protocols.
} 
\label{fig:results:perf}
 \vspace{-2mm}
\end{figure}

\noindent Figure~\ref{fig:results:perf} shows the  execution time of our workloads when we run them with the five different protocols. 
We normalize execution time to the {\em WB} configuration, which has
high performance but is intolerant to faults.

We observe that $WT$ is prohibitively expensive, especially for write-intensive workloads such as  Ocean-cp and Ocean-ncp. Writing through remote writes
can be very slow in TSO, as  writes from the same core cannot overlap. As a result, the SB
can fill-up,
which results in the  stall of the
core execution. The average execution time of $WT$ over $WB$ is 
3.68x.

\sysnamex-baseline serializes the Coherence
and Replication transactions for a remote store. We see that it is still 
significantly slower than $WB$. On average, its execution time is 
2.88x higher than $WB$.
\sysnamex-baseline  penalizes write-intensive applications by slowing down  SB  draining. 
However, because of its write-back approach, it is a faster and resilient alternative to  $WT$.

\sysparallel   overlaps the Coherence and the Replication transactions for a remote store. 
However, we observe that  the benefits over \sysnamex-baseline  are minor (only 3\% on average). The reason  is that the architecture we model is equipped with 
exclusive prefetching. 
As soon as a store's target address is known, a request for ownership is sent to the memory system. Thus, when the store reaches the head of the SB, the Coherence transaction has often completed, exposing the Replication transaction like  in \sysnamex-baseline.  

\sysproactive 
proactively initiates the Replication transaction for a remote store as soon
as the store enters the SB. 
The advantage  of this technique is twofold. First,  
it overlaps the Replication transaction with the store's waiting time in the SB.
Second, it {\em enables the overlapping of the Replication transactions for the stores in the SB}. 
As a result, this design substantially boosts performance. 
The execution time with \sysproactive is on average only
30\% higher than $WB$, while making the protocol resilient to failures. 

\begin{figure}[t]
    \vspace{-2mm}
    \centering
    \includegraphics[width=.80\linewidth]{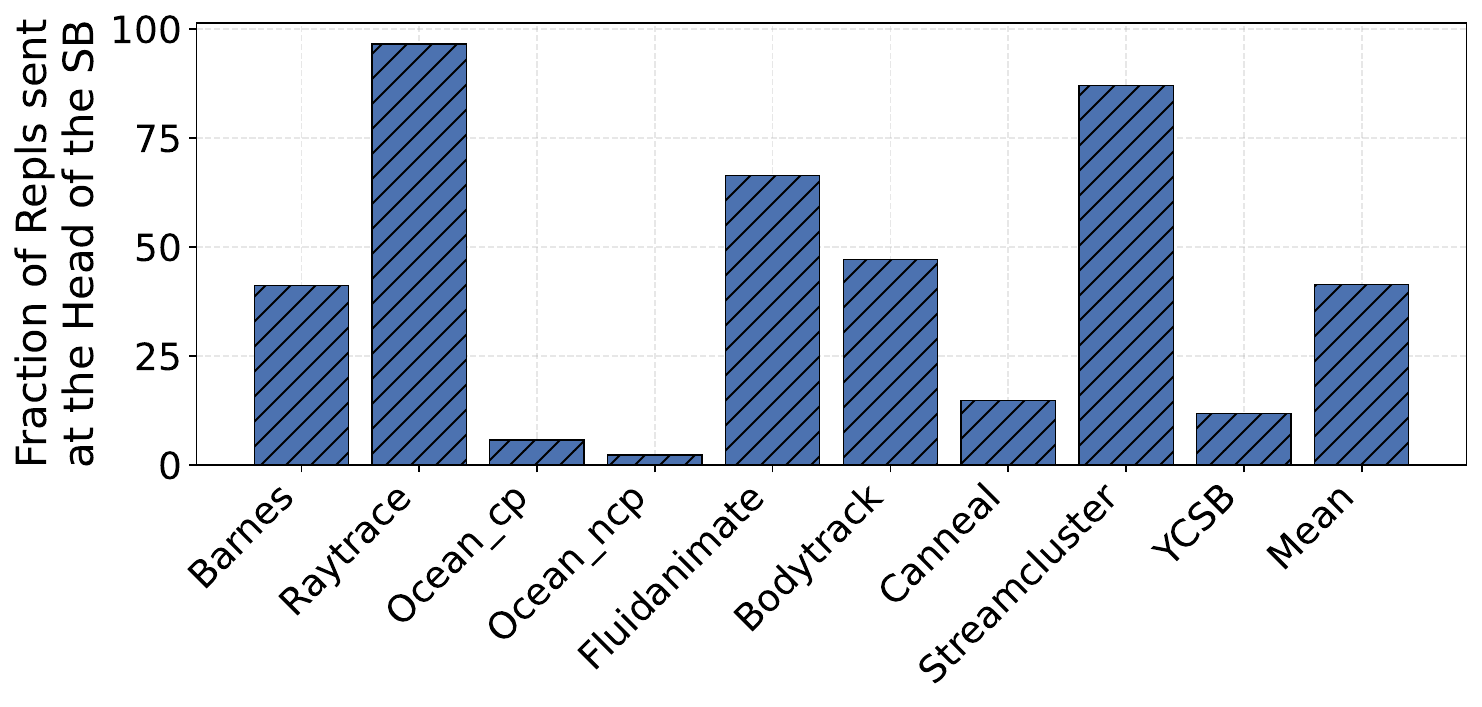}
    \vspace{-2mm}
    \caption{Fraction of \texttt{REPL}s sent by \sysproactive when the store is at the head of the SB.}
    \label{fig:results:sbheadReplfraction}
\end{figure}

Figure~\ref{fig:results:perf} shows that  raytrace and fluidanimate     do not
benefit much from \sysproactivex. To understand why, Figure~\ref{fig:results:sbheadReplfraction}
shows the fraction of \sysproactive  \texttt{REPL}s that are sent when the store is at the head of 
the SB. This is the latest time that \sysproactive can send the \texttt{REPL}s. If it sends them
at this time, there is no difference between \sysparallel and \sysproactivex. The higher this
fraction is, the lower   the
difference between  \sysproactive and \sysparallel is. We see that raytrace and fluidanimate
have a high fraction, which explains their behavior in Figure~\ref{fig:results:perf}. In
these applications, a store often remains in the SB without sending  \texttt{REPL}s, with the
hope of coalescing with a future store that never appears.

Streamcluster also has a high fraction in Figure~\ref{fig:results:sbheadReplfraction}. However,
because the fraction of requests that are remote writes is relatively small, all protocols
behave similarly (Figure~\ref{fig:results:perf}). \vspace{-2mm}

\subsection{Characterization Analysis}\label{sec:eval:characterization}

We now analyze several aspects of \sysproactivex.

\begin{figure}[t]
    \vspace{-2mm}
    \centering
    \includegraphics[width=1\linewidth]{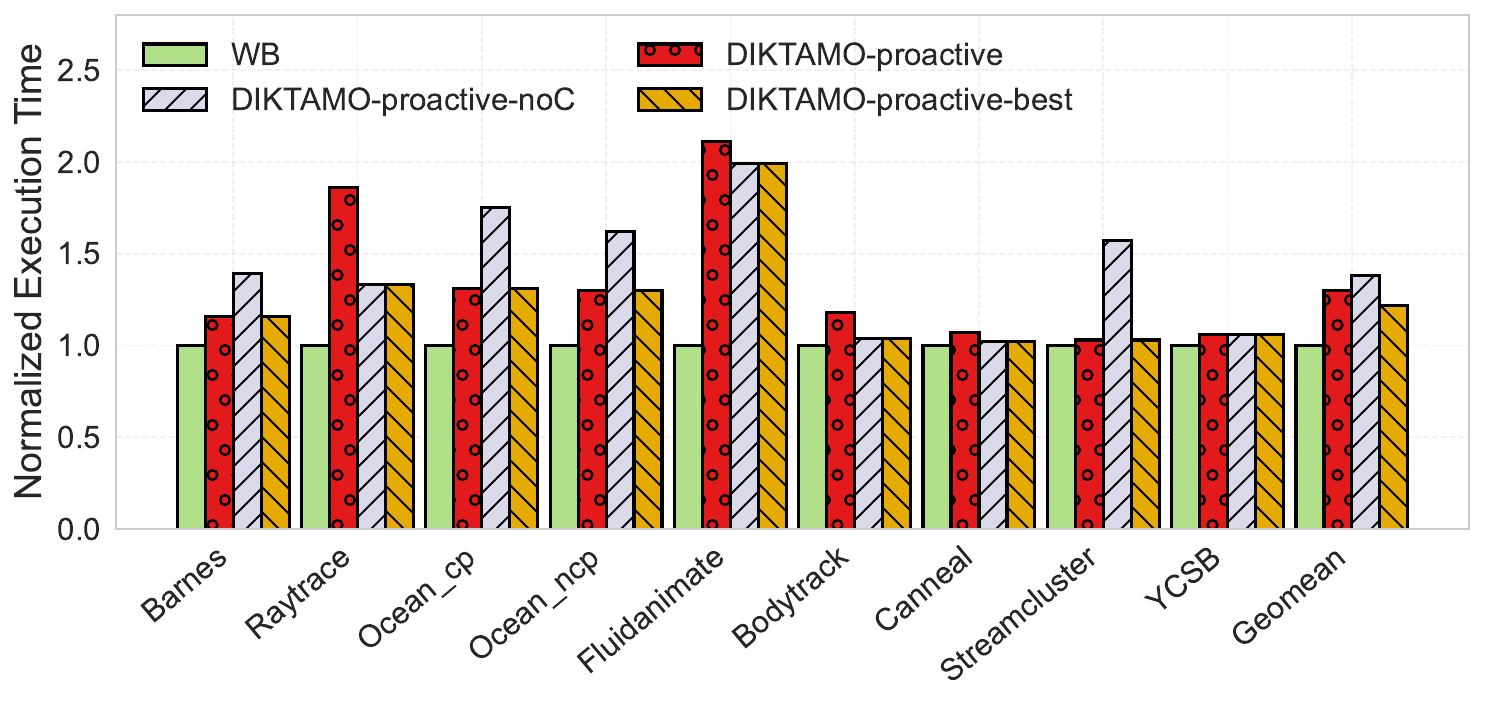}
    \vspace{-4mm}
    \caption{Evaluating the impact of store coalescing on \sysproactivex.}
    \label{fig:results:coalescing}
\end{figure}

\vspace{1mm}
\noindent\textbf{Impact of Store Coalescing on \sysproactivex.} As discussed in Section~\ref{sec:diktamo:replication}.5, supporting store coalescing in \sysproactive
required a design that sometimes delays sending \texttt{REPL}s for stores in the SB. Hence,
supporting coalescing affects performance in two ways.
On the one hand, it can improve performance by combining multiple
stores  and  reducing the number of write transactions. On the other hand,
it may delay sending \texttt{REPL}s for some writes, 
reducing the ability to hide the Replication transaction.

Figure~\ref{fig:results:coalescing} examines whether the positive or the negative effects of 
supporting coalescing dominates. It repeats Figure~\ref{fig:results:perf} for 
$WB$ and \sysproactive (which supports coalescing), and adds two more bars: 
\sysproactivex-noC (which does not support coalescing) and \sysproactivex-best, which is 
simply the best of \sysproactive and \sysproactivex-noC.

Comparing \sysproactive with \sysproactivex-noC shows that \sysproactive is only slightly
better, and that applications vary widely.
Some applications like ocean-cp benefit  from supporting coalescing, while others like raytrace lose performance. In general, it is hard to correlate Figure~\ref{fig:results:coalescing}
with Figure~\ref{fig:results:sbheadReplfraction}: Figure~\ref{fig:results:sbheadReplfraction} simply shows whether the \texttt{REPL}s were delayed to
their latest times; it does not show when the \texttt{REPL}s were sent or
whether fruitful coalescing occurred. Still, we see that
raytrace, which delayed many \texttt{REPL}s in Figure~\ref{fig:results:sbheadReplfraction}, ends up being hurt
by
supporting coalescing in Figure~\ref{fig:results:coalescing}. Finally, the lower execution time of
\sysproactivex-best
suggests that selectively enabling coalescing on and off may be interesting.
\sysproactivex-best's average overhead over $WB$ is only 22\%, while \sysproactivex's is
30\%.

\begin{figure}[t]
    \vspace{-2mm}
    \centering
    \begin{subfigure}[b]{0.85\linewidth}
        \centering
        \includegraphics[width=1\linewidth]{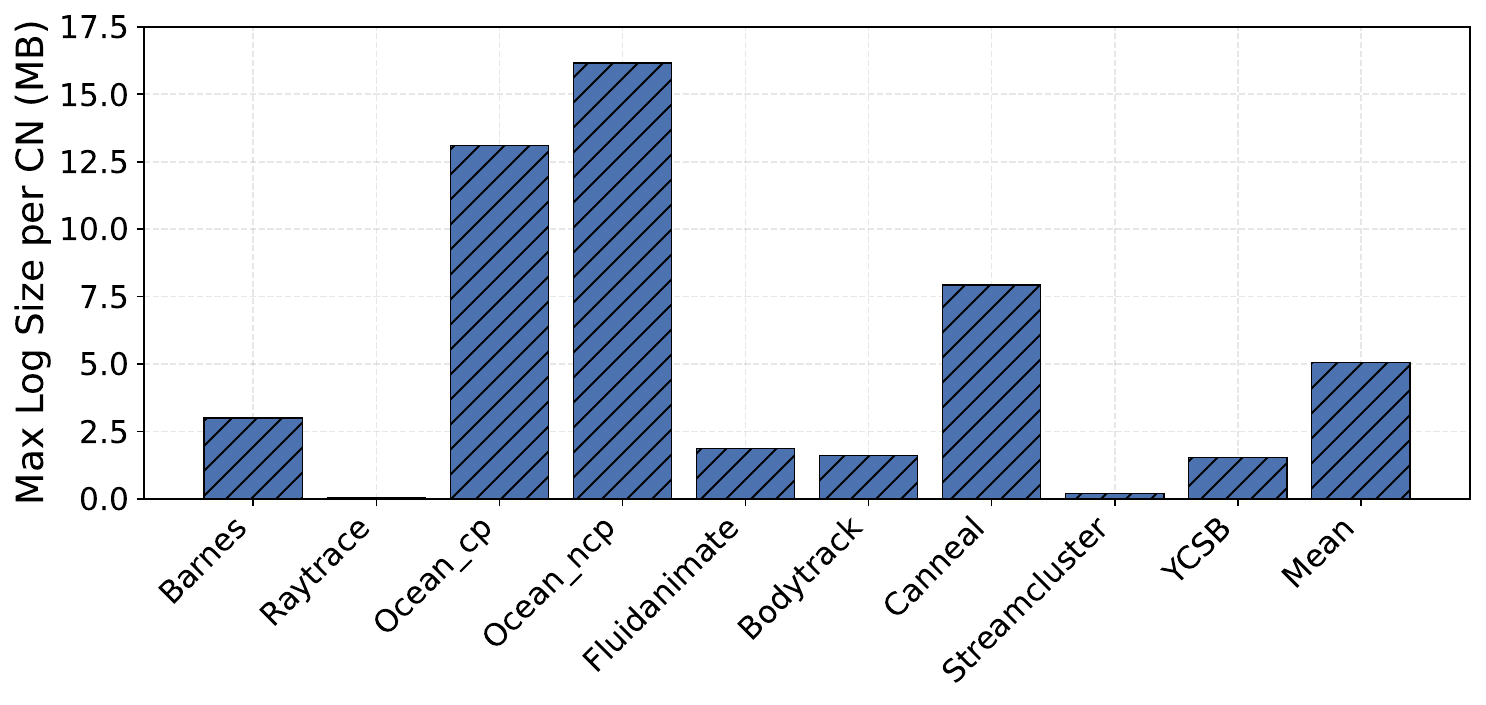} 
    \end{subfigure}
    \caption{Maximum uncompressed log size generated in any CN  in \sysproactive  between consecutive log dumps.}
    \label{fig:sensAnalysis:maxLogCN}
\end{figure}

\vspace{1mm}
\noindent\textbf{Log Size.} Each Logging Unit has a 4KB SRAM Log Buffer for pending transactions and
an 18MB DRAM Log Buffer to log updates. 
Figure~\hbox{\ref{fig:sensAnalysis:maxLogCN}} shows the maximum 
uncompressed log size 
generated in any CN  
in \sysproactive between consecutive log dumps. This is the biggest possible DRAM buffer size needed. In practice, a much smaller space is needed, since  the data is being
compressed as it is being deposited into the DRAM buffer, before being
flushed to MNs at the end of the period between log dumps. However, to be   conservative,
we size the
DRAM buffer in each Logging Unit to be 18MB.

\begin{figure}[t]
    \vspace{-3mm}
    \centering
    \begin{subfigure}[b]{0.90\linewidth}
        \centering
        \includegraphics[width=1\linewidth]{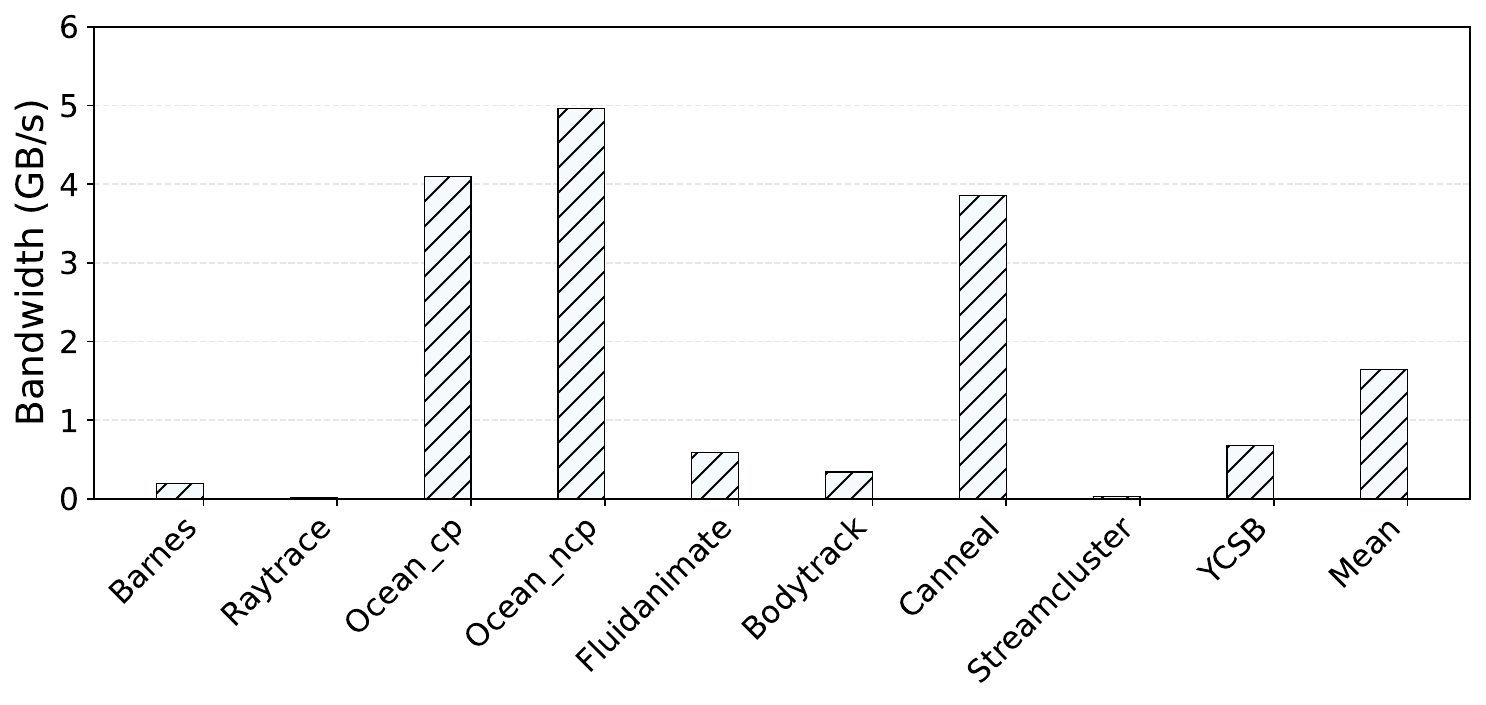}
    \end{subfigure}
    \caption{Average CXL network bandwidth 
    consumed by the 16 CNs periodically saving the compressed logs 
    in \sysproactivex.}
    \label{fig:sensAnalysis:bwConsumption}
\end{figure}

\vspace{1mm}
\noindent\textbf{CXL Network Bandwidth Consumed by Saving the Logs.}
Figure~\ref{fig:sensAnalysis:bwConsumption} shows the average CXL network bandwidth consumed by the 16 CNs periodically saving the compressed logs to the 
MNs in \sysproactivex.
We see that 
the bandwidth consumption 
is minimal and not higher than
5\,GB/s 
across all the workloads.
Hence, the execution time of workloads is  largely unaffected by the log saving operations. 

\vspace{1mm}
\noindent\textbf{Lines Owned by a Crashed Node.}
As part of the recovery, a handler traverses the directory and finds the addresses of
all the lines that were Shared or
Owned by the caches in the crashed CN. Owned lines can be Dirty or Exclusive in caches, but the directory does
not distinguish between them. For the Owned lines, the handler then traverses the logs to recover
their value before the crash. 

\begin{figure}[t]
    \centering
    \vspace{-2mm}
    \begin{subfigure}[b]{0.95\linewidth}
        \centering
        \includegraphics[width=1\linewidth]{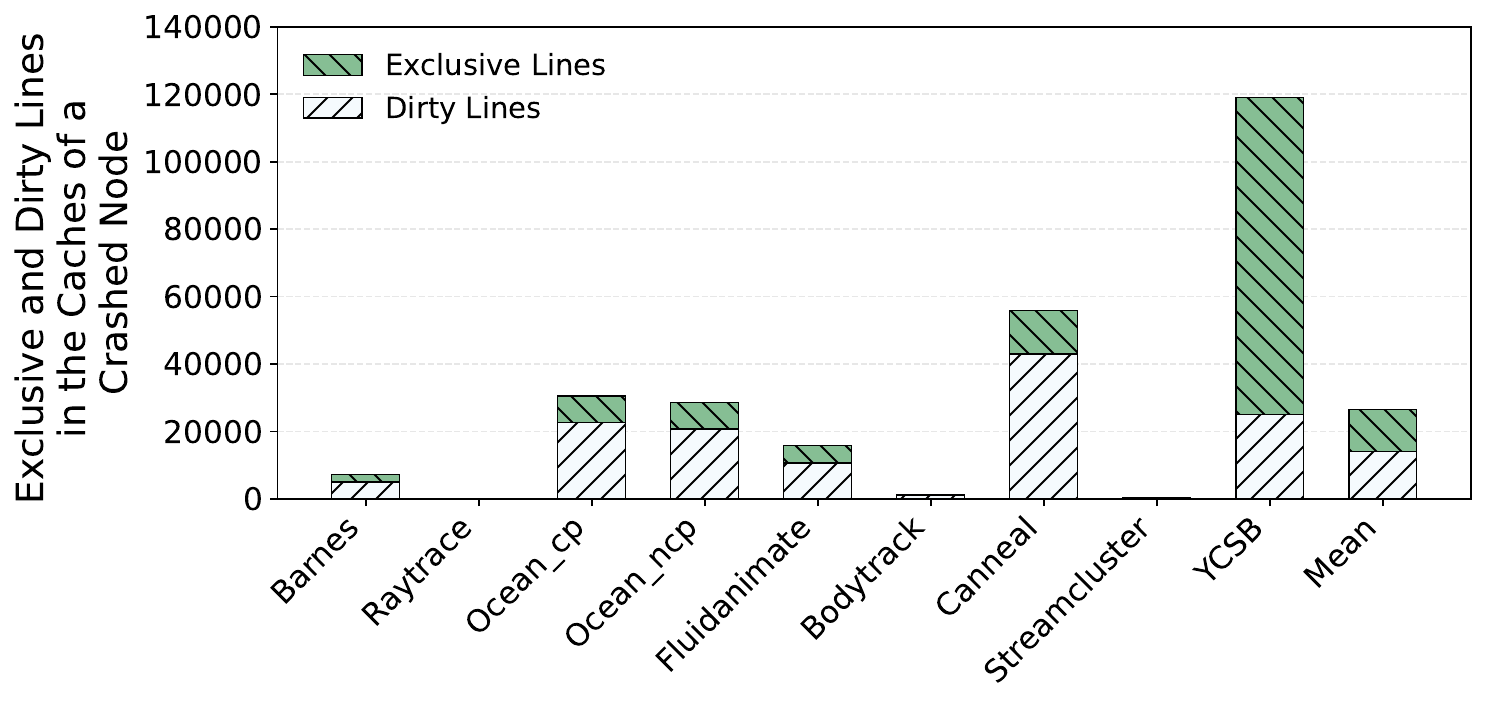} 
    \end{subfigure}
    \vspace{-2mm}
    \caption{Number of Exclusive  and Dirty lines in the caches of a crashed CN in \hbox{\sysproactivex}.}
    \label{fig:recovery:dirty-lines}
\end{figure}

In this experiment, we trigger the crash of CN 0 at 12.5ms of execution in the applications under \sysproactivex, and count the number of Owned lines. The directory gives us a number,
and then we check the caches of the crashed CN to count those that are indeed Dirty.
We label the difference between the two counts as Exclusive, although some of them  may have been evicted silently from the caches. 
Figure~\ref{fig:recovery:dirty-lines} shows the result for the applications. For reference, the maximum 
 number of different lines in the caches of a CN is 163K. 
The figure shows that, on average, less than  29K
lines are Owned, although, in YCSB, such number reaches 119K.

\begin{figure}[t]
    \centering

    \begin{subfigure}[b]{0.42\textwidth}
        \centering
        \includegraphics[width=\linewidth]{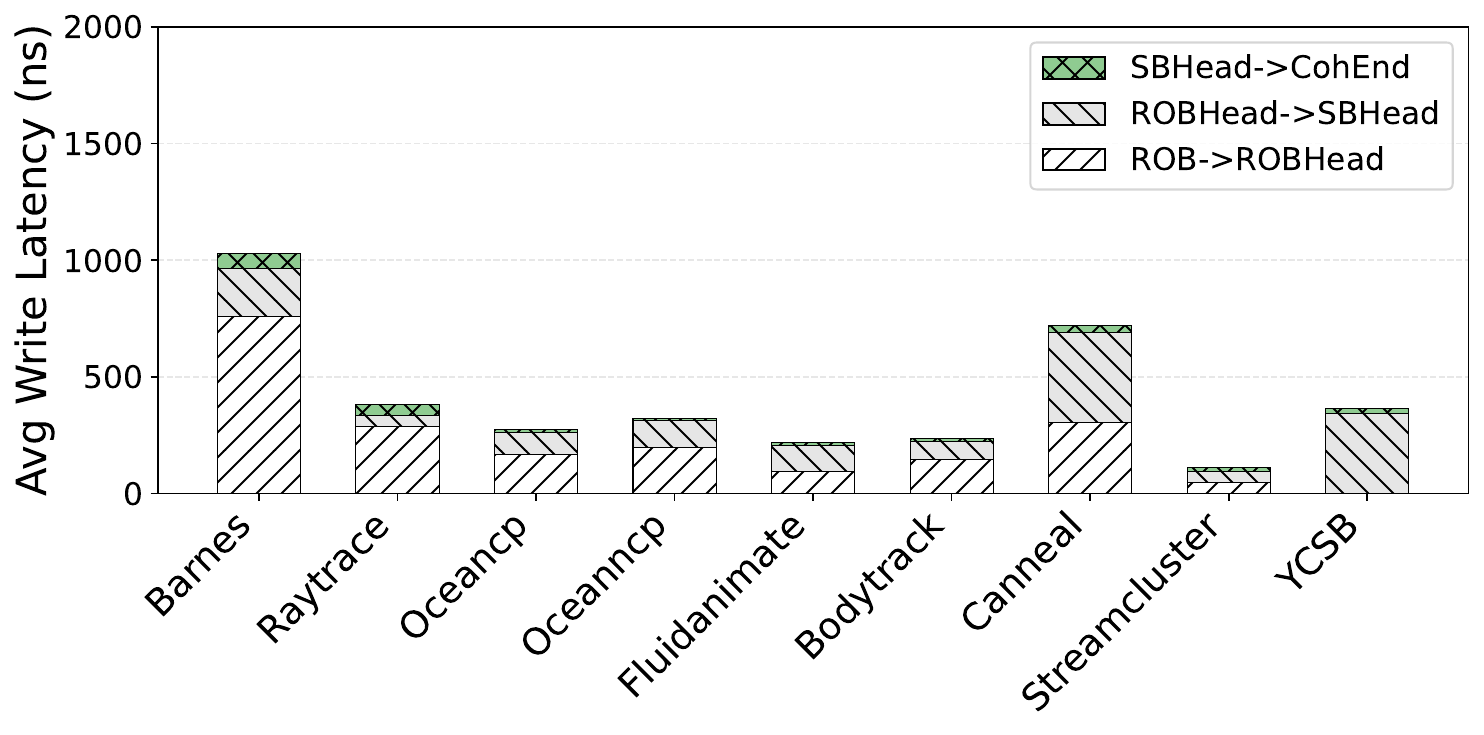}
        \vspace{-7mm}
        \caption{Write Back (WB)}
        \label{fig:remWRbkdown_wb}
    \end{subfigure}
    \hfill
    \begin{subfigure}[b]{0.42\textwidth}
        \centering
        \includegraphics[width=\linewidth]{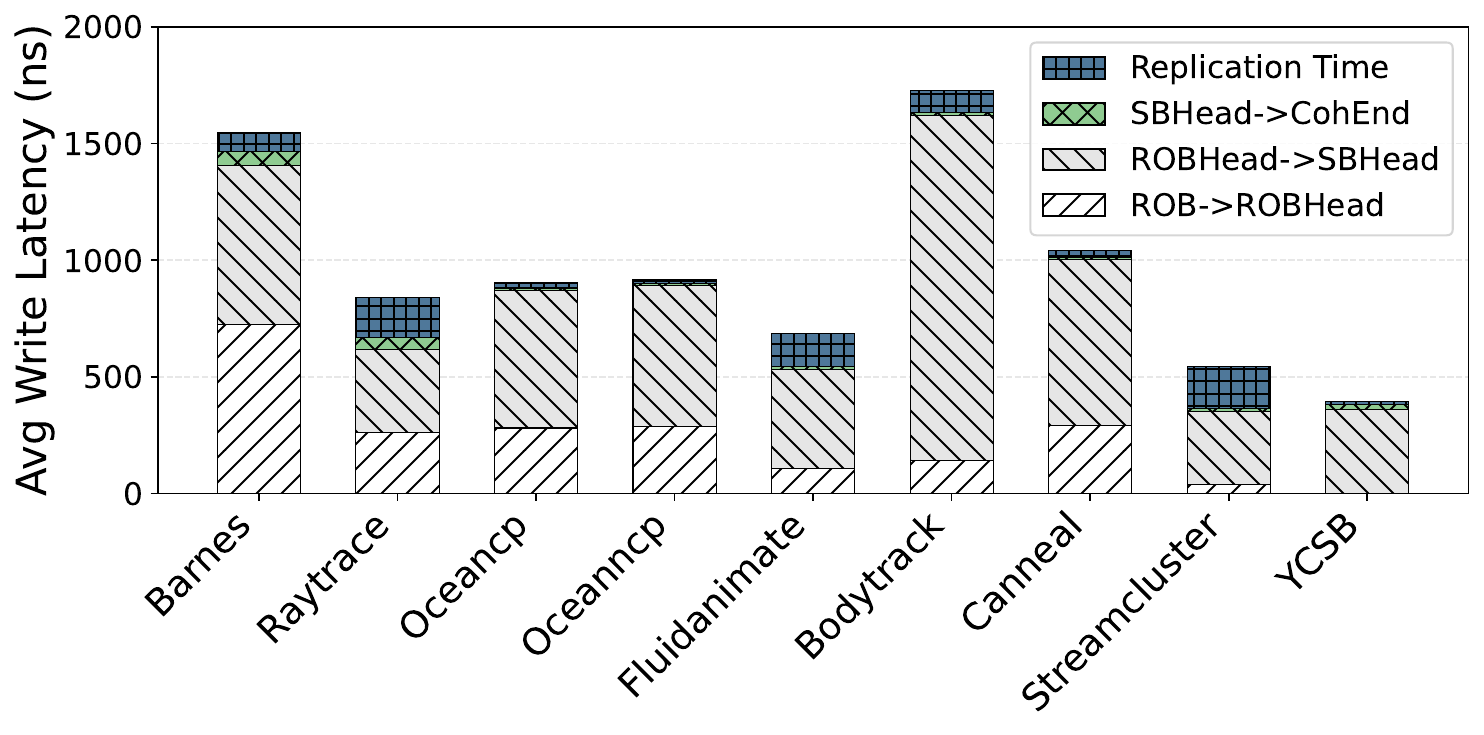}
        \vspace{-7mm}
        \caption{\sysproactivex{}}
        \label{fig:remWRbkdown_pro}
    \end{subfigure}
    \hfill
    \begin{subfigure}[b]{0.42\textwidth}
        \centering
        \includegraphics[width=\linewidth]{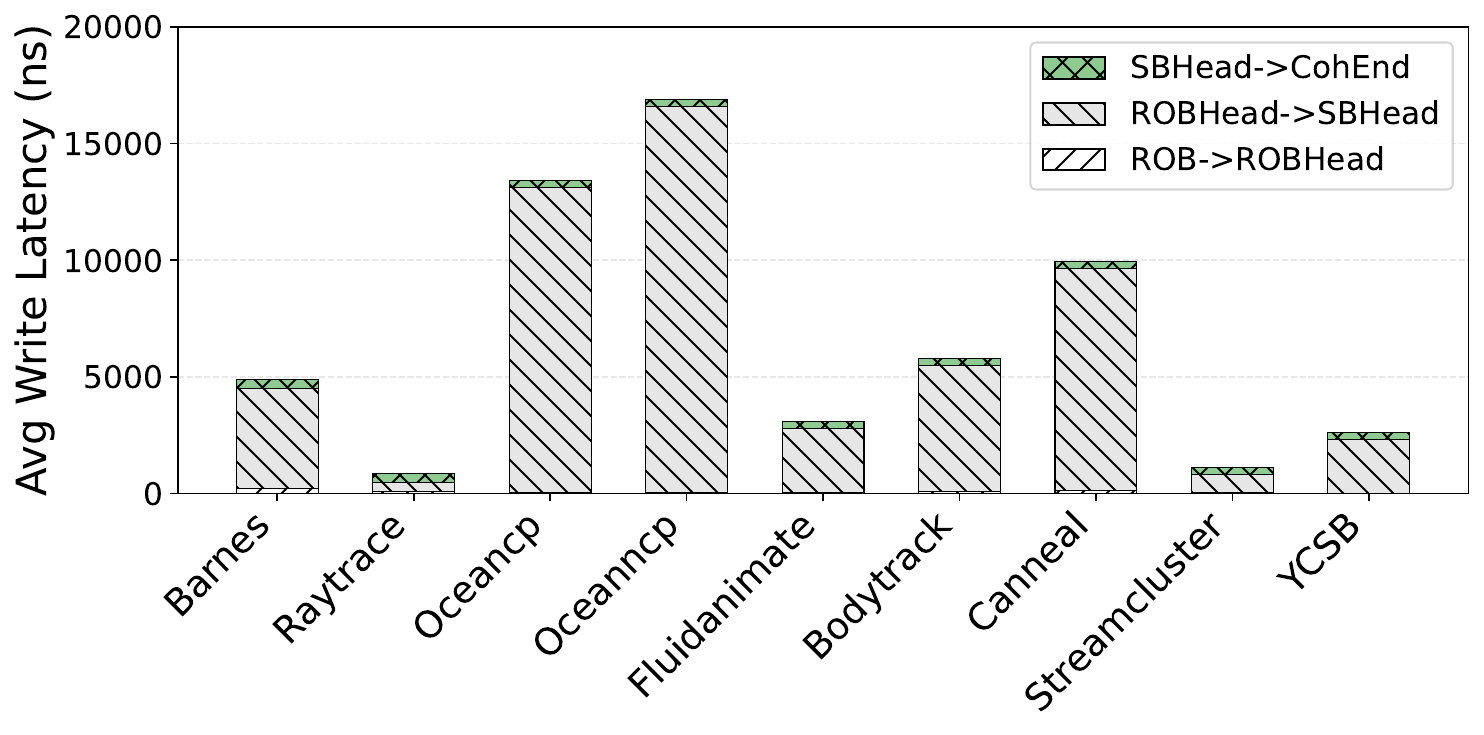}
        \vspace{-7mm}
        \caption{Write Through (WT)}
        \label{fig:remWRbkdown_wt}
    \end{subfigure}
    
    \caption{Breakdown of the average latency of a write to remote data 
     for three different protocols. Note that the Y axis of the WT subplot (c) spans a range 10$\times$ larger than the other subplots.}
    \label{fig:remWRbkdown}
    \vspace{-1mm}
\end{figure}

\vspace{1mm}
\noindent\textbf{Remote Write Latency Breakdown.}
Figure~\hbox{\ref{fig:remWRbkdown}} shows the average latency of a write to
remote data broken 
down into its components:
\hbox{\textit{ROB$\rightarrow$ROBHead}} is the time from when the store is deposited into the ROB until it retires;
\hbox{\textit{ROBHead$\rightarrow$SBHead}} is the time from store retirement
until the store reaches the head of the SB.
\hbox{\textit{SBHead$\rightarrow$CohEnd}} is the additional time it takes for the coherence transaction to complete, bringing the line in state Exclusive to L1;  and
\hbox{\textit{Replication Time}} is any additional time beyond
CohEnd that the write needs to remain in the SB waiting for the Replication transaction to complete.
After \textit{SBHead$\rightarrow$CohEnd} and \textit{Replication Time}, the write commits.
Results are shown for $WB$ (a), \textit{\sysproactivex} (b),  and $WT$ (c).
 
We see that,  
in $WB$ (Figure~\ref{fig:remWRbkdown_wb}), \textit{ROB$\rightarrow$ROBHead} and
\textit{ROBHead$\rightarrow$SBHead} dominate. \textit{SBHead$\rightarrow$CohEnd}
is minimized by exclusive prefetching and \textit{Replication Time} does not
exist.
Moving to \textit{\sysproactivex} (Figure~\ref{fig:remWRbkdown_pro}), 
\textit{ROB$\rightarrow$ROBHead} 
remains  mostly the same, while
\textit{ROBHead$\rightarrow$SBHead} increases in all cases
and becomes dominant. 
The average write spends more time in the SB
because of the increased number of operations performed in a write transaction.
In addition,  
\textit{Replication Time} appears in
several applications. 
The applications that have the highest increase in average remote write latency
are bodytrack and streamcluster.
 
Correlating these write latency increases with the increases in execution time in Figure~\ref{fig:results:perf} is not
easy, since not all the remote write latency is visible in the execution's critical path,
and there are many factors that affect execution time.
What is clear is the correlation between the \textit{Replication Time} in
Figure~\ref{fig:remWRbkdown_pro}
and the fraction of \texttt{REPL}s sent at the SB head shown in Figure~\ref{fig:results:sbheadReplfraction}.
Applications with a
high fraction of \texttt{REPL}s sent at the SB head in Figure~\ref{fig:results:sbheadReplfraction}
(raytrace, streamcluster and, to a lesser extent,
fluidanimate, barnes, and bodytrack) have a 
high \textit{Replication Time} in
Figure~\ref{fig:remWRbkdown_pro}.

Figure~\hbox{\ref{fig:remWRbkdown_wt}} shows the 
data for \hbox{$WT$}. 
Note that the Y axis in Figure~\hbox{\ref{fig:remWRbkdown_wt}} spans a range 10$\times$ larger than the other subplots.
Unlike \textit{\hbox{\sysproactivex}}, \hbox{$WT$} has no replication component. On the other hand, \hbox{\textit{ROBHead$\rightarrow$SBHead}}  entirely dominates the latency. This is because, under TSO, WT stores often wait substantially for 
older stores to complete their round trip to the MN before draining from the SB. Across applications, we see a pattern like in Figure~\hbox{\ref{fig:sensAnalysis:maxLogCN}}, which shows that applications with heavier remote-store activity also tend to require larger \hbox{\sysnamex} logs. Overall, the figure shows why \hbox{$WT$} is costly: it turns remote-write intensity into long SB waiting time, whereas \textit{\hbox{\sysproactivex}} keeps WB caching and overlaps replication with existing SB delays.

\subsection{Recovery Latency}
\label{sec:eval:recovery}

This section evaluates the recovery latency of 
\hbox{\sysnamex}. Based on 
Section~\hbox{\ref{sec:failure-handling}}, we break down the 
operations during recovery into the seven steps
in Table~\hbox{\ref{table_rec}}.
\texttt{RecBegin} and \texttt{RecEnd} capture the synchronization overhead required to first start and finally end
the recovery. 
\texttt{CM$\leftrightarrow$MN} is the exchange of recovery-control messages between the Configuration Manager (CM) and the  MN directory controllers. 
\texttt{TravDir} is the traversal of the directory state to identify cache lines affected by the failed CN. 
\texttt{MN$\leftrightarrow$LU} is the exchange of information between MN directory controllers and Logging Units (LUs). 
\texttt{TravLU} is the search for candidate replicas of the affected cache lines in the LUs. 
Finally, \texttt{MNMerge} 
is identifying the correct version of each lost cache line (either from
those received from LUs or from searching the logs already in MNs) and
saving them to memory.
The recovery is the same for all the \hbox{\sysnamex} flavors. In this experiment, we crash one node after 12.5ms of execution and recover the system.

\begin{table}[t]
    \centering
    \caption{Steps of the recovery in \hbox{\sysnamex}.}
    \resizebox{.99\linewidth}{!}{
        \begin{tabular}{|l|p{6.8cm}|}
            \hline
            \textbf{Step} & \textbf{Description} \\
            \hline
            \texttt{RecBegin} &
            Upon detection of the failure, the CM interrupts the surviving CNs, and waits until they quiesce, complete pending transactions, and acknowledge. \\
            \hline
            \texttt{CM$\leftrightarrow$MN} &
            The CM sends recovery-start messages to the MN directory controllers. Once the MN directory controllers complete their recovery, they send acknowledgments to the CM. \\
            \hline
            \texttt{TravDir} &
            Each MN directory controller scans its directory to identify entries for cache lines shared or owned by the failed CN. For each shared line, remove the CN from the sharer list. For each owned line, save the address of the line in a \enquote{failed-CN-owned lines} list.\\
            \hline
            \texttt{MN$\leftrightarrow$LU} &
            The MN directory controllers send the \enquote{failed-CN-owned lines} list to the Logging Units. Later, the Logging Units return the latest versions of the lines in the \enquote{failed-CN-owned lines} list (if they had them). \\
            \hline
            \texttt{TravLU} &
            Each Logging Unit scans its log for validated entries matching the lines in the \enquote{failed-CN-owned lines} list and compiles a list with the most recent version of each line that it can provide. \\
            \hline
            \texttt{MNMerge} &
            Each MN directory controller identifies the correct versions of the lost cache lines (either from those received from LUs or from searching the logs already in MNs) and saves them to memory. The directory entries are updated accordingly.\\
            \hline
            \texttt{RecEnd} &
            The CM broadcasts recovery completion to the surviving CNs  and waits for acknowledgments. \\
            \hline
        \end{tabular}
    }
    \label{table_rec}
\end{table}

\begin{figure}[t]
    \centering
    \vspace{-5mm}
    \includegraphics[width=.99\linewidth]{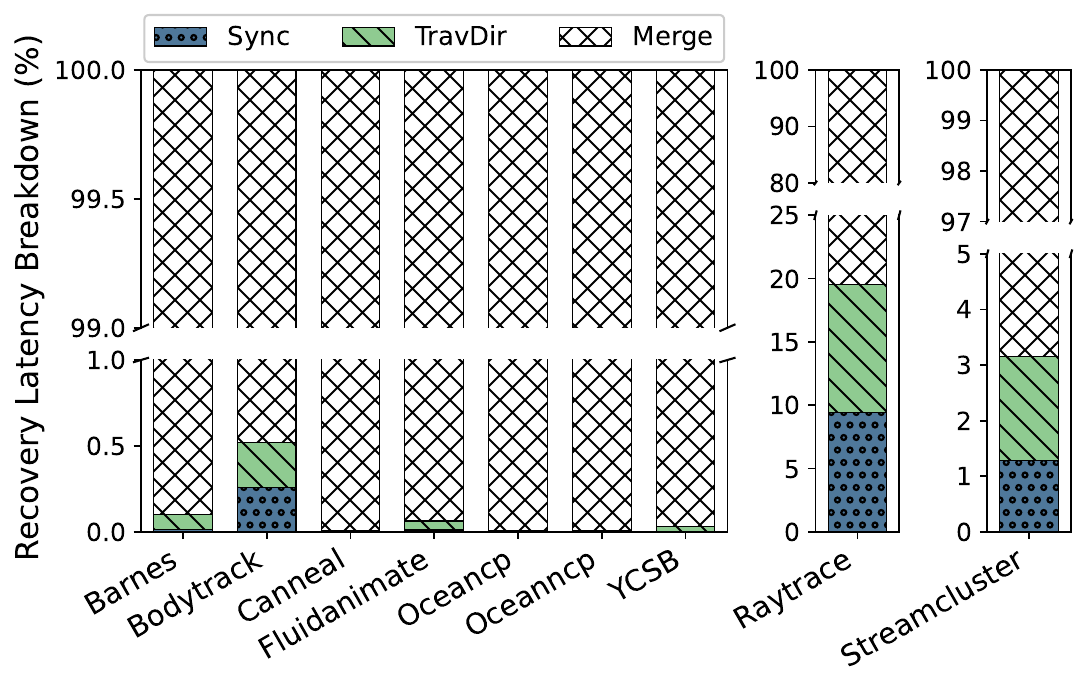}
    \vspace{-6mm}
    \caption{Normalized latency of the recovery in \hbox{\sysname} broken down into three main components.}
    \label{fig_rec}
    \vspace{-1mm}
\end{figure}

Figure~\hbox{\ref{fig_rec}} shows the normalized latency of the steps
of the recovery. We show three separate plots because of the different behavior of the applications. 
In each plot, the Y-axis is broken into two pieces to ensure all categories are visible. 
We are interested in comparing three components. The first one is the time
for the CNs, MNs, and LUs to synchronize and communicate. We call it \texttt{Sync} and 
includes \texttt{RecBegin}, \texttt{CM$\leftrightarrow$MN}, \texttt{MN$\leftrightarrow$LU}, and \texttt{RecEnd}.
The second component is the   traversal   of the directories. This is the 
\texttt{TravDir} time. The third component is finding the lost cache line
versions 
by searching the logs in the Logging Units or in the MNs, and merging
them to memory. We call it \texttt{Merge} and includes \texttt{TravLU} and \texttt{MNMerge}.
From the figure, we see that,
in most cases, the very large 
fraction of the time is spent in \texttt{Merge}---finding the lost versions
and merging them to
memory.
This is costly because it involves substantial searching. In comparison,
traversing the directory (\texttt{TravDir}) and communicating/synchronizing
(\texttt{Sync}) take much less time. \texttt{TravDir} is sped-up by sequential
prefetching. 

\subsection{Sensitivity Analyses}
\label{sec:eval:sensitivity}

In this section, we perform two sensitivity analyses.

\begin{figure}[t]
    \centering
    \begin{subfigure}[b]{.96\linewidth}
        \centering
            \includegraphics[width=1\linewidth]{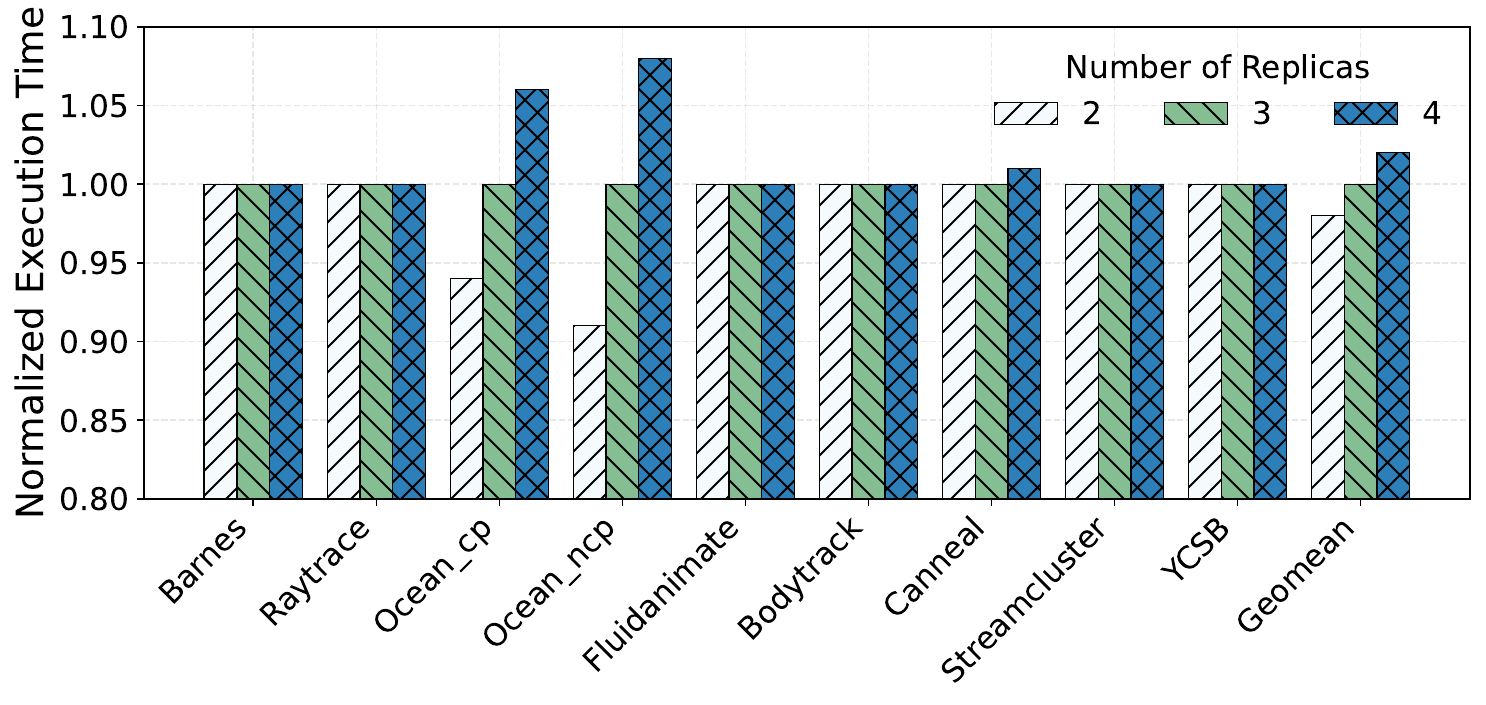} 
    \end{subfigure}
    \caption{Execution time with \sysproactivex{} varying the replication factor $N_r$. All bars are normalized to $N_r$=3.
    }
    \label{fig:sensAnalysis:replFactor}
\end{figure}

\vspace{1mm}
\noindent\textbf{Replication Factor.}
Figure~\ref{fig:sensAnalysis:replFactor} shows the execution time of the applications with \sysproactive for 
replication factors $N_r$ equal to 2, 3, and 4. All bars are normalized to the default $N_r$=3.
On average, the execution time increases   with $N_r$.
A system with $N_r$=4 takes   2\% longer to execute than one with $N_r$=3, while one with
$N_r$=2 is 2\% faster than it. Remote write–intensive
applications like  ocean-cp  and  ocean-ncp  are significantly affected.
From this data, it can be shown that
\sysproactive  with $N_r$=2 takes 27\% longer than $WB$ to execute our workloads.

\begin{figure}[t]
    \centering
    \vspace{-1mm}
    \begin{subfigure}[b]{\linewidth}
       \includegraphics[width=1\linewidth]{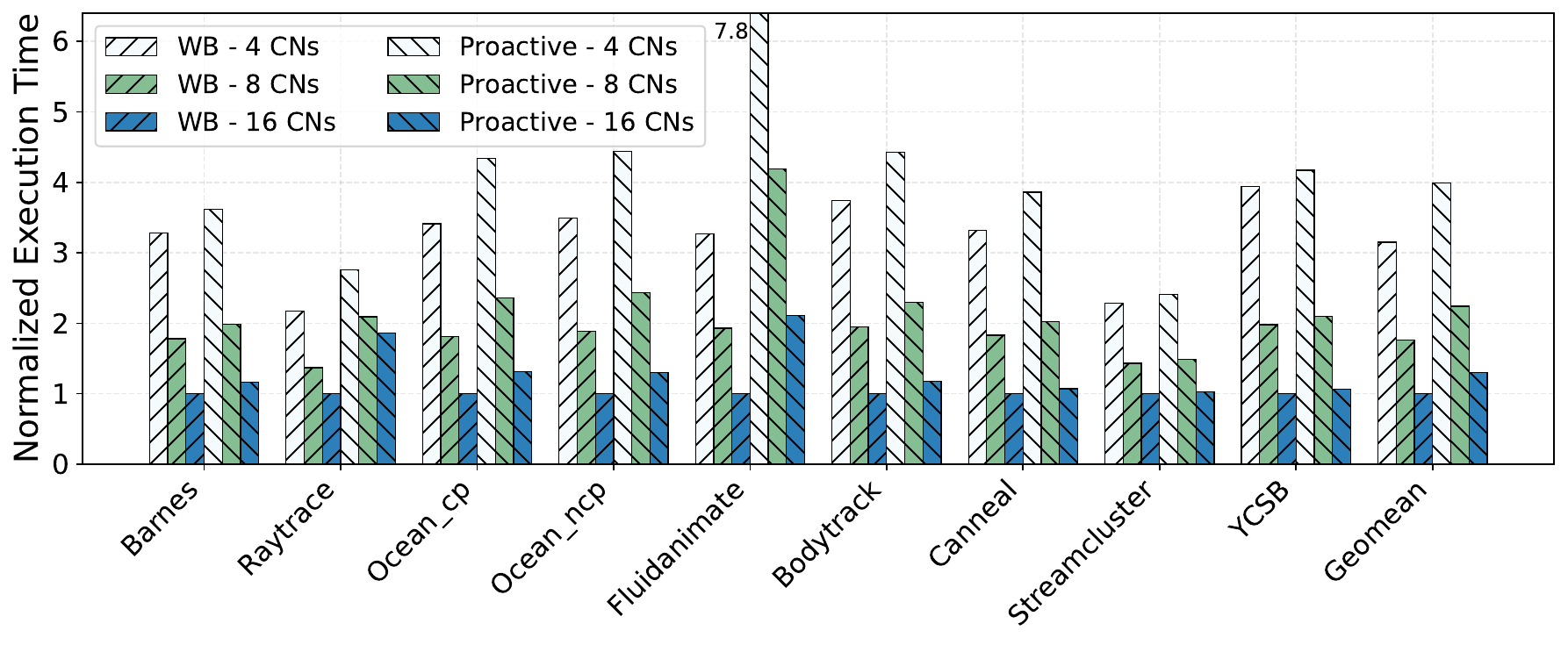} 
    \end{subfigure}
    \caption{Execution time of \sysproactivex{} and {\em WB} for different numbers of CNs. Bars are normalized to  WB with 16 CNs.} 
    \label{fig:sensAnalysis:numCNs}
\end{figure}

\vspace{1mm}
\noindent\textbf{Scalability with the Number of CNs.}
Past work suggests that CXL systems with 16 CNs are a good design point~\cite{pond-asplos23, hydrarpc-atc24},
and some CXL switches support 32 ports~\cite{XConn_XC50256} (which is ideal for our 16 CN  and 16 MN system). Figure~\ref{fig:sensAnalysis:numCNs} shows the execution time of \sysproactive and {\em WB} as we change the number of CNs from 4 to 16. The bars are normalized to $WB$ for 16 CNs.
The figure shows that, across all applications, \sysproactive has good scalability.
On average, going from 4 to 16 CNs, reduces the execution time by 3.15x and 3.07x for {\em WB} and \sysproactivex, respectively.

%% file: Sections/relatedWork.tex
\section{Related Work}\label{sec:relatedWork}

\noindent
{\bf Related uses of CXL.} Several systems have leveraged CXL for memory expansion and pooling. CXL-ANNS~\cite{cxlanns-atc23} uses CXL memory to scale approximate nearest neighbor workloads.
Memstrata~\cite{memstrata-osdi24} introduces a memory allocator in the hypervisor and the first hardware-managed memory tiering for CXL. Pond~\cite{pond-asplos23} addresses memory stranding issues by enabling flexible pooling.

With CXL 3.0, memory sharing has enabled new distributed system abstractions. HydraRPC~\cite{hydrarpc-atc24} implements low-latency RPC over shared CXL memory. CXLfork~\cite{cxlfork-asplos25} enables fast process forking over CXL shared memory. TrEnv~\cite{trev-sosp24} explores the use of shared memory over CXL to accelerate serverless functions. 

Tigon~\cite{tigon-osdi25} and CTXNL~\cite{ctxnl-asplos25} are transactional systems
over CXL. They both use software cache coherence for the majority of the data in the program and CXL-supported hardware cache coherence for metadata. Both works provide 
little or no discussion on recovery. Tigon uses
conventional database transaction logging and recovers by concurrently restoring the log. CTXNL says it \enquote{relies on the transaction processing frameworks to
ensure data consistency}. Hence, 
recovery in these systems is software-based and coarse-grained. 

Apta~\cite{apta-dsn23} examines a  CXL environment where nodes can fail. 
It introduces a write-through coherence protocol for fault
tolerance in Function-as-a-Service (FaaS) systems. It is a hardware-supported \emph{transactional} protocol.
Each transaction reads an object, operates on it, and uses
special hardware to atomically push the updated object to the main memory node
(there is a  single  main-memory node). It supports linearizability and,
therefore, conflicting transactions are squashed. The protocol
introduces lazy-invalidation, which moves invalidations out of
the critical path of a write, so that the writer is not blocked if a
sharer fails. Special scheduling support ensures lazy invalidations
do not introduce inconsistencies. 

In contrast to these three works, \sysname targets general-purpose (not just transactional or application-specific)
environments. Unlike Apta, it uses a write back protocol,
does not require
special
scheduling, and works in a cluster with multiple memory nodes (not just one).
Unlike
all the schemes, it uses a 
novel hardware-supported, fine-grained recovery scheme for performance.

\noindent
{\bf Resilience through replication and coherence protocols.}
Replication has long been used to achieve both resilience
and performance in distributed and shared-memory systems. 
Works such as MINOS~\cite{minos-hpca24}, FARM~\cite{farm-nsdi14}, PAR~\cite{par-atc19}, and RAMCloud~\cite{ramcloud-sosp11}  replicate  data across multiple nodes for high availability and low-latency access.
Other work explores cache-line replication for fault tolerance in processor and memory systems. ICR~\cite{zhang-dsn03} replicates active L1 lines by overwriting predicted-dead blocks. 
Replication Cache~\cite{zhang-tc05} extends this with a small replica cache storing dirty-line copies on each write. Fernandez-Pascual~et~al.~\cite{fernandez-hpca07} ensure at least one valid copy by delaying invalidations until safe transfer, tolerating interconnect faults. Dvé~\cite{patil-isca21} replicates DRAM cache lines across NUMA sockets to provide resilience to memory errors, and maintains coherence among the replicas, but  does not handle host-wide failures. In contrast, \sysnamex{} logs updates
in the context of CXL, enabling recovery from node crashes. 

Shared-data reliability and coherence fault-tolerance has been studied 
for shared-memory multiprocessors~\cite{hive,fernandez-hpca07,rsm}.  
Our work complements these efforts by providing high-performance fault tolerance in 
distributed CXL systems. 

%% file: Sections/conclusion.tex
\section{Conclusion}\label{sec:conclusion}

To address limitations of the CXL 
specification, this paper extends CXL to make it resilient to node failures.
We call the system  {\em \sysnamex}.
\sysname augments the
coherence transaction of writes to create replicas in 
the Logging Units of a few other nodes.
Recovery after a node failure involves using the logs 
to bring the directory and memory to a correct state.
Our evaluation with 16 4-core nodes shows that \sysname 
enables fault-tolerant execution
with   30\% slowdown with three replicas (or 27\% with two replicas)
over the same platform with no fault-tolerance support. Further, \sysname is 
2.82x 
faster than
ensuring reliability by using
write-through writes.